\documentclass[aip,preprint,floatfix]{revtex4-1}
\usepackage{amsfonts,amssymb,amsmath,array}

\usepackage{dcolumn}
\usepackage{bm}

\usepackage[final]{graphicx}
\usepackage{graphicx}
\usepackage{epstopdf}
\graphicspath{{./Figures/}} 

\usepackage[usenames]{color} 

\usepackage{soul} 

\begin{document}

\title{Transport of active particles in an open-wedge channel}

\author{Lorenzo Caprini}
\affiliation{Gran Sasso Science Institute (GSSI), 
Via F.Crispi 7, I-67100 L'Aquila, Italy}

\author{Fabio Cecconi}
\affiliation{Istituto dei Sistemi Complessi (CNR), Via Taurini 19, I-00185 Roma, Italy}

\author{Umberto Marini Bettolo Marconi}
\affiliation{Scuola di Scienze e Tecnologie, Universit\`a di Camerino,  
Via Madonna delle Carceri, I-62032, Camerino, Italy}

\begin{abstract}
The transport of independent active Brownian particles within a 
two-dimensional narrow channel, modeled as an open-wedge, is studied 
both numerically and theoretically. 
We show that the active force tends to localize the particles near the 
walls thus reducing the effect of the entropic force which, instead, is 
prevailing in the case of passive particles.
As a consequence, the exit of active particles from the smaller side of the 
channel is facilitated with respect to their passive counterpart.
By continuously re-injecting particles in the middle of the wedge, we obtain 
a steady regime whose properties are investigated with and without the 
presence of an external constant driving field.
We characterize the statistics and properties of the exit process from the 
two opposite sides of the channel, also by making a comparison between the 
active and passive case.
Our study reveals the existence of an optimal value of the persistence time 
of the active force which is able to guarantee the maximal efficiency in the 
transport process. 
\end{abstract}

\maketitle

\section{Introduction}

In the last years, the theoretical study of self-propelled microswimmers has become an 
important research area at the crossroads between biology, mathematics, and physics.
These systems are ubiquitous in nature,
typical examples being bacteria \cite{berg2008coli}, protozoa \cite{blake1974mechanics}, spermatozoa \cite{woolley2003motility} and 
living tissues \cite{poujade2007collective} and actin filaments \cite{kohler2011structure} to mention just a few. 
On the other hand, bioengineers and physicists are developing techniques to create artificial
self-propelled objects, as in the case
of the so-called Janus particles \cite{walther2013janus,lattuada2011synthesis}. 

The common feature of active particles is the existence of a self-propelling mechanism 
converting the environmental energy into motion 
\cite{bechinger2016active,romanczuk2012active,marchetti2013hydrodynamics,ramaswamy2010mechanics}. 
The nature of such propulsion varies from one 
microswimmer to another, but in general determines a ballistic motion at short spatial and temporal scales
and a diffusive motion at larger scales.

In order to describe the behavior of such systems,
different theoretical models have been developed, among them we recall
a) the Run\&Tumble  (R\&T) model \cite{nash2010run,tailleur2008statistical,sevilla2014theory}, where the microswimmers alternatively perform at a given rate
ballistic displacements and tumbles, i.e.  random changes of the orientation of their velocity.  b)
a continuous model with a Langevin-like dynamics, the so-called Active Brownian particle (ABP) model 
\cite{ten2011brownian,romanczuk2012active}
described in detail in Sec.\ref{SecModel}. c) the active Ornstein-Uhlenbeck particle (AOUP) model 
\cite{szamel2014self,das2018confined,marconi2015towards}.
All these models show an intriguing phenomenology ranging
from particle accumulation near the walls of a container 
\cite{MinoClement2018EColi,caprini2019active,wensink2008aggregation,kaiser2012capture,
fily2014dynamics,elgeti2015run,Elgeti_2013} 
to the existence of non-Boltzmann probability
distributions \cite{fodor2016far,marconi2017heat,caprini2018linear} also in the presence 
of acoustic traps \cite{takatori2016acoustic,caprini2019activity}, 
the appearance of negative mobility in the presence of non convex potentials \cite{caprini2018activeescape} and motility 
induced phase separation (MIPS) 
\cite{fily2012athermal,buttinoni2013dynamical,bialke2015active,cates2015motility,
speck2016collective,tjhung2018cluster,digregorio2018full}, in the case of interacting active particles.
The influence of geometrical constraints on the motion of active particles is a less explored and only partly understood issue,
in spite of its importance in elucidating how some systems of biological 
interest behave.
For instance, living bodies harbor colonies of bacteria normally localized 
in the skin, external mucosae, gastrointestinal tracts etc. 
These bacteria, by crossing narrow constrictions, are able to invade/infect 
the hosts' internal tissues that instead need to remain sterile 
\cite{ribet2015bacterial}. How this passage occurs is a problem of great 
relevance for evident reasons, especially in cases of pathogen infections.

In the framework of R\&T modeling, first-passage properties have been studied both with \cite{malakar2018steady} 
and without \cite{weiss2002some,angelani2014first} thermal noise for a one-dimensional channel, whereas the same 
problem was numerically studied for a one-dimensional version of the ABP-model \cite{scacchi2018mean} and in two-dimensional
corrugated channels in Refs.~\cite{malgaretti17,cinesi}.
In a similar context, a Fick-Jacobs transport equation \cite{jacobsBook} 
accounting for the channel geometry via an entropic effective force 
\cite{zwanzig,regueraFJ,burada_Rev} has been proposed 
for weakly active particles \cite{dagdug14}. 

In the present paper, we idealize the motion of bacteria by means of ABP and model the pore as a narrow wedge-shaped capillary. 
At variance with previous works, we investigate how the first-passage process 
of an ABP through a narrow constriction depends on the
activity parameters as well as the geometry. 
The distribution of the escape events has been addressed by several 
authors \cite{caprini2018activeescape,sevilla2019stationary,fily2018self,sharma2017escape}, in this 
work we focus on the general features of escape process, 
in particular comparing the efficiency of ``active'' transport with the Brownian transport.  

The paper is organized as follows: in Sec.\ref{SecModel} we introduce the model, while in Sec.\ref{Sec3} 
the steady-state properties of active 
particles in the channel are discussed, including the density along the transport direction and 
the density along a section of the pore.
In Sec.\ref{SecEscape}, we study the escape-time statistics
showing how the efficiency of the active transport depends on the active force. 
Finally, we summarize the main results in the conclusive section.

\section{Model of active particles in open wedge-geometry}
\label{SecModel}
\begin{figure}[t!]
\includegraphics[clip=true,keepaspectratio,width=0.9\columnwidth]
{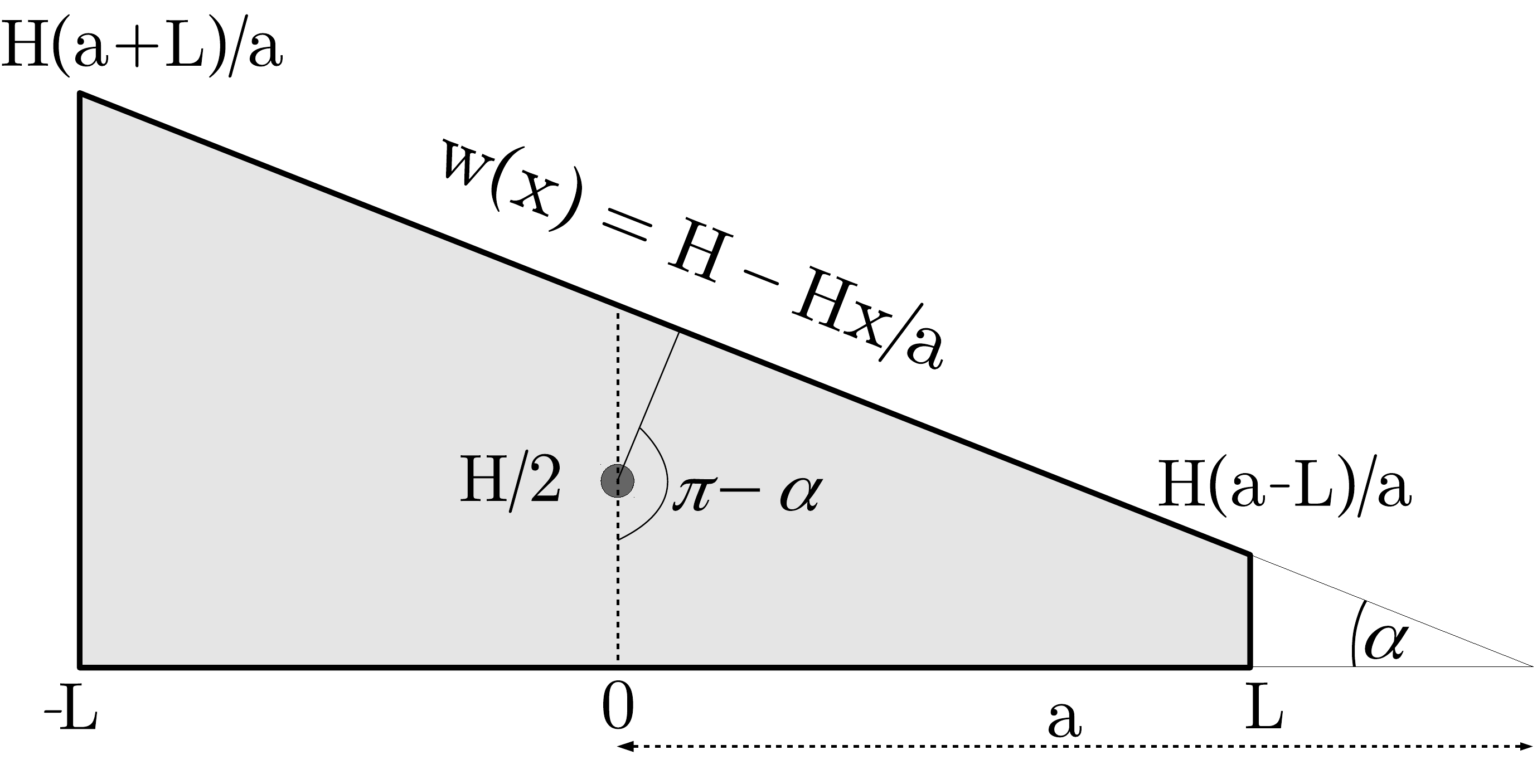}
\caption{\label{fig:wedge} Sketch of the truncated wedge used as 
simulation box $Q =\{|x|\le L, 0\leq y\le w(x)\}$, 
bounded by the lines $y = H - H x/a$ and $y=0$.
Absorbing conditions are placed at $x=\pm L$, for which the 
particles crossing $x=\pm L$ are removed from the system.  
The black circle, centered in $(0,H/2)$, marks 
the narrow region where particles are either initially emitted 
or re-injected after their absorption.}
\end{figure}
We consider an assembly of independent active particles immersed in a viscous solvent and constrained to move 
in the two dimensional truncated-wedge channel shown in Fig.\ref{fig:wedge}. We neglect the inertial effect and
consider the over-damped dynamics of the particles, where the position of
 each particle, $\mathbf{r}$, moves according to the following 
 stochastic differential equation
\begin{equation}\label{eq:motion}
\gamma\dot{\boldsymbol{r}} =\boldsymbol{F}({\mathbf r}) + 
\epsilon {\mathbf {\hat x}}+ \gamma U_0 {\mathbf {\hat e}}(t)\;,
\end{equation}
where $\boldsymbol{F}$ is an external force, $\epsilon {\mathbf {\hat x}}$ 
a drift along the channel axis representing a systematic bias associated to a drag 
or a biological bias towards positive $x$.
The constant $\gamma$ is the friction coefficient.
The last term of Eq.\eqref{eq:motion} represents the ABP self-propulsion mechanism, namely a 
force of fixed strength, $\gamma U_0$, and varying orientation, 
${\mathbf {\hat e}}(t) = (\cos\theta(t),\sin\theta(t))$, whose 
angle $\theta(t)$ evolves
according to the following Wiener process:
\begin{equation}\label{eq:motion2}
	\dot{\theta}=\sqrt{2 D_r} \xi ,
\end{equation}
where the constant $D_r$ is the rotational diffusion coefficient and $\xi$ a 
white noise with zero average and unitary variance.
As several experimental studies indicate \cite{bechinger2013physics}, 
the influence of the thermal agitation of the solvent surrounding the 
microswimmers can be neglected \cite{romanczuk2011brownian}. 

The particles are confined to the domain 
$Q =\{(x,y)\;:\;|x|\le L, 0\leq y<w(x)\}$ 
bounded by the bottom of the open-wedge channel at $y = 0$ and by its
upper boundary
\begin{equation}
w(x) = \dfrac{H}{a}(a - x) .
\end{equation}
The left and right vertical boundaries, at  $x=\pm L$, are absorbing, while 
both boundaries are soft reflecting walls (no-flux boundaries).
Moreover, we are interested in the narrow channel condition: $H \ll L$.
The top wall exerts on the particles a force 
directed along its normal direction $\mathbf n = (w'(x),-1)/\sqrt{1 + w'(x)^2}$, 
whereas the repulsion of the bottom wall is directed along $\hat{\mathbf{y}} = (0,1)$.
To represent this force we introduce a wall-potential
 $V(u) = V_0/m (\sigma/u)^m$, where $V_0$ defines its energy scale and $\sigma$ its length-scale
 assumed to be small with respect to $H,L$ and write:
\begin{equation}
\mathbf F = -V'(w(x)-y)\;\mathbf{n} - V'(y)\; \hat{\mathbf{y}}
\label{forceu}
\end{equation}
where the prime represents the derivative with respect to the argument $u$.
The form of the force~\eqref{forceu} determines specular reflection when particles ``collide'' with 
the walls.
To prevent excessive penetration, the functional form of $V$ must guarantee
strong repulsion when evaluated at $y = w(x)$ and $y=0$ and thus we assume 
$\sigma$ to be at least $\sim 10^{-2} H$ and $V_0=1$, $m=4$.

\begin{figure*}[!t]
\includegraphics[clip=true,width=0.8\textwidth,keepaspectratio]
{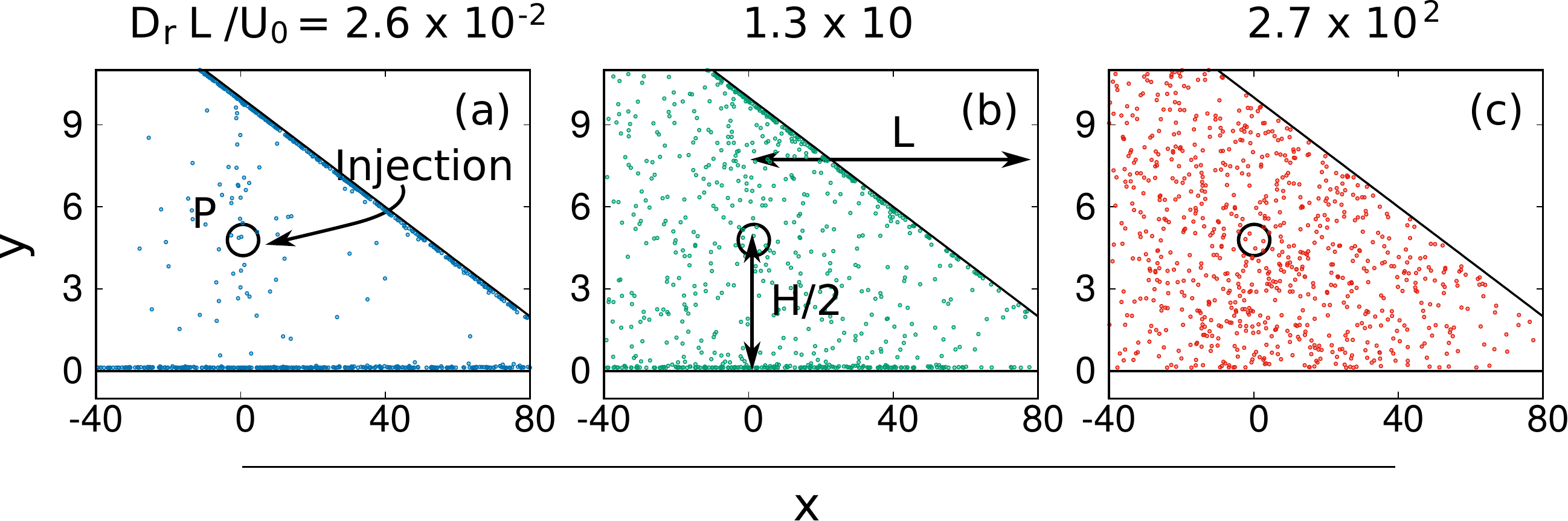}
\caption{
Snapshots of particle configurations, upon re-injection at the site $(0, H/2)$, for different values of the parameter $D_r L/U_0$
in the absence of field $\epsilon = 0$. Panels (a), (b) and (c), showing a different degree of particle accumulation to the walls, are obtained with
$D_r L /U_0=2.6 \times 10^{-2}, 1.3 \times 10, 2.7 \times 10^2$, respectively. Other system parameters are $L=80$, $a=100$, $H=10$, and $U_0=3$.
\label{fig:Snapshotfree}}
\end{figure*}

\begin{figure}[!t]
\includegraphics[clip=true,width=0.8\columnwidth,keepaspectratio] 
{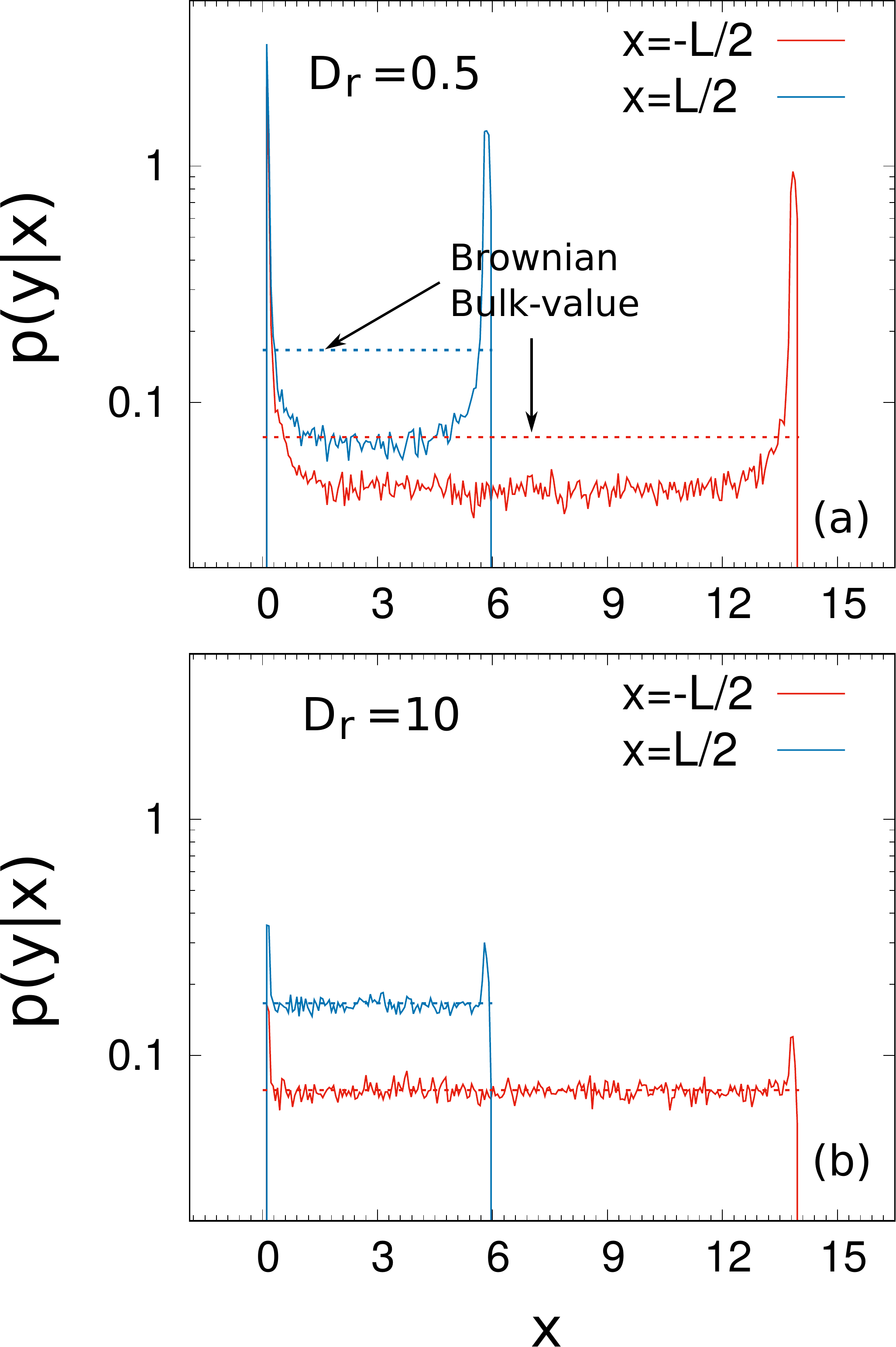}
\caption{Conditional probability distribution functions, $p(y|x)$, 
evaluated at $x = L/2$ (blue line) and $x=-L/2$ (red line). 
Panel a) and b) are obtained for $D_rL/U_0=1.3 \times 10$, $2.7 \times 10^2$, 
respectively.  
System parameters are: $L=80$, $a=100$, $H=10$, $\epsilon=0$, $U_0=3$.
}\label{fig:pcond_Y}
\end{figure}

\begin{figure}[!t]
\includegraphics[clip=true,width=0.9\columnwidth,keepaspectratio] 
{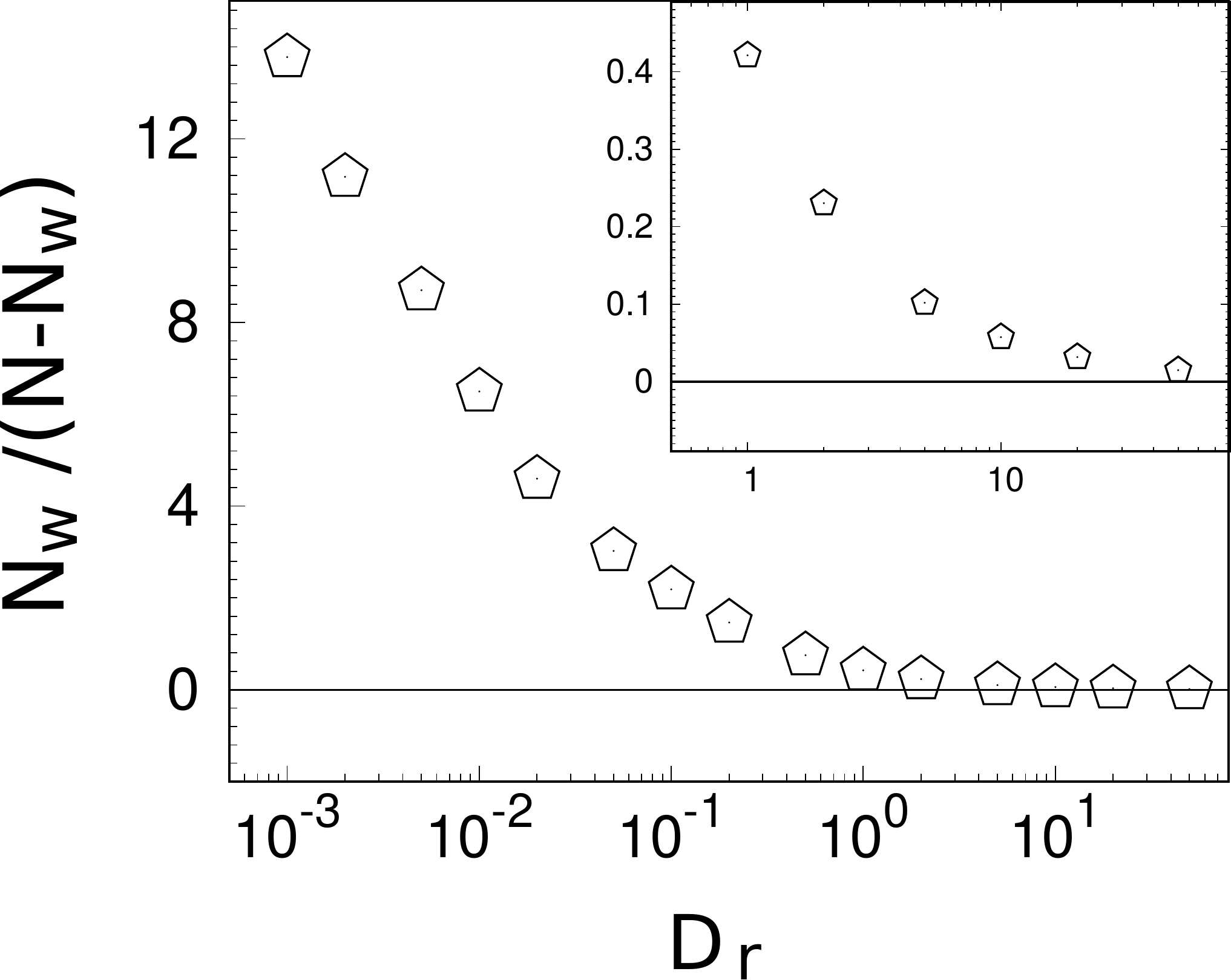}
\caption{Ratio $N_w/(N-N_w)$, between the number of particles accumulating 
at the walls and the remaining ones in the bulk, as a function of $D_r$. 
The inset is a blow-up of the range $D_r > 1$.
System parameters are the same as in Fig.\ref{fig:pcond_Y}.
}\label{fig:fig3}
\end{figure}

\begin{figure*}[!t]
\includegraphics[clip=true,width=0.8\textwidth,keepaspectratio]
{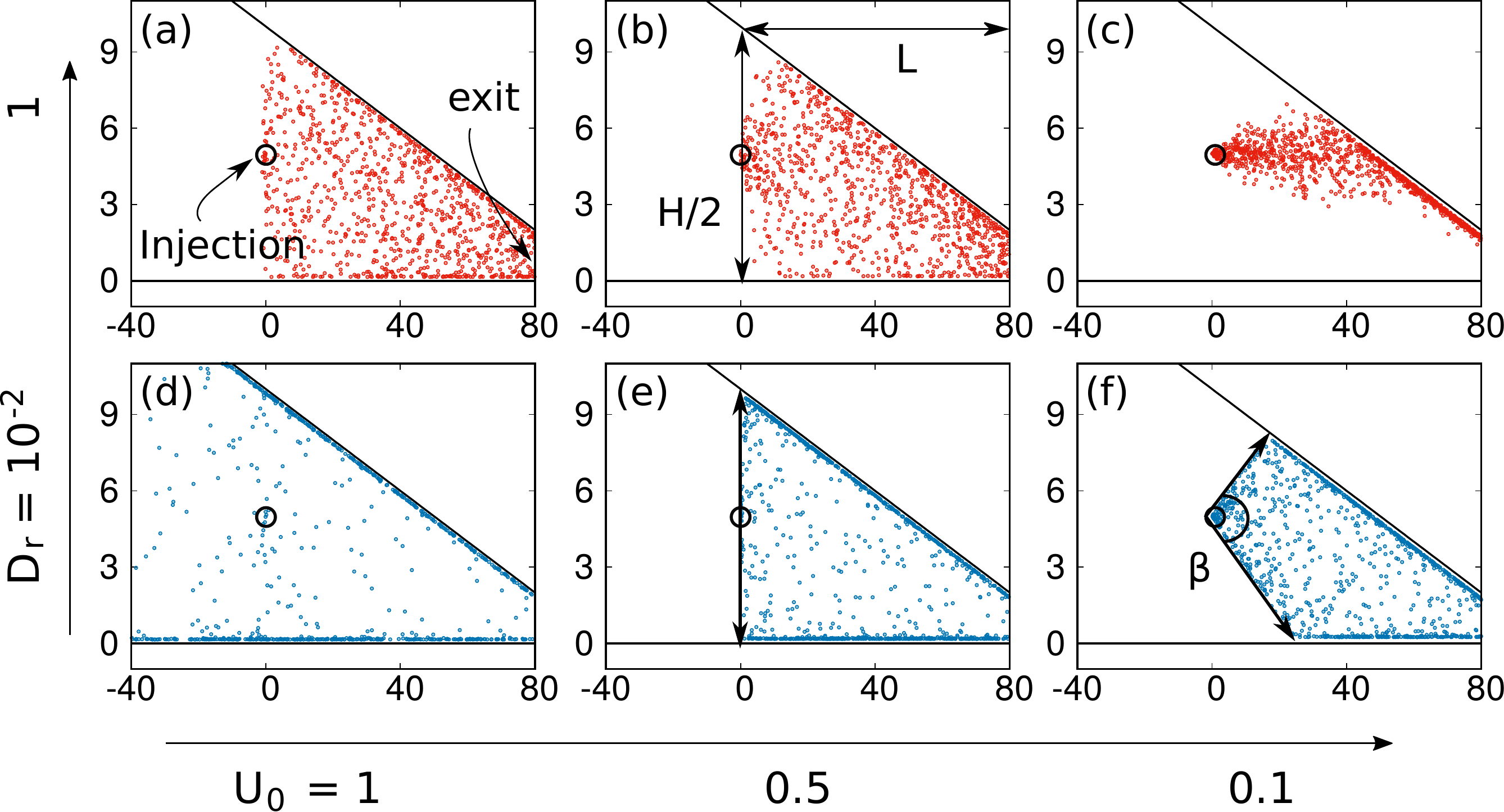}
\caption{Snapshots of particle positions in the presence of a field $\epsilon = 0.5$ at different $U_0$ and $D_r$. Panels (a), (b) and (c) refer to values $U_0 = 1.0, 0.5, 0.1$
and the same $D_r = 1.0$, while panels (d), (e) and (f) are obtained with the same set of $U_0$ but a value $D_r = 0.01$. Black lines represent the walls, and the gray circle
centered at $(0, 5)$ is the area where particles are reinjected when they cross the right exit at $x=L=80$. The remaining parameters are $a=100$ and $H=10$.}
\label{fig:Snapshotforce}
\end{figure*}

In the numerical simulations,
the particles are initially placed in a small neighborhood of the point 
$P=(0,H/2)$ (see Fig.\ref{fig:wedge}), mimicking the injection by means of a 
``micro-pipette'', and eventually leave the pore at 
the $L$ and $-L$ boundaries. A stationary process is achieved by reinserting
these particles at the point $P$.
Correspondingly, the direction of the active force acting on the re-injected 
particles is obtained from a uniform distribution of angles $\theta$ in the 
interval $[0,2\pi]$.
The evolution of $N=10^4$ particles is obtained by integrating 
Eq.\eqref{eq:motion} with a Euler-Maruyana algorithm 
\cite{toral2014stochastic}, at least up to time $\mathcal{T} \sim 10^3 /D_r$. 
Throughout the paper, the geometry will be fixed such that $H=10$, $L=80$ 
and $a=100$, which guarantees the condition $H\ll L$.

We begin our analysis by first discussing how the particle distribution 
over the domain $Q$ influences the escape properties when $D_r$ is varied.

\subsection{Case $\epsilon = 0$}
In Fig.\ref{fig:Snapshotfree} we display three snapshots of particle 
configuration at increasing values of $D_r L/U_0$, in the absence of an 
external force.

Panels a) and b) clearly show thin denser stripes near both the 
upper and lower wall, 
indicating the tendency of strongly active particles to ``climb on'' 
confining edges.
With the growth of $D_r L / U_0$, the accumulation at 
walls decreases from a) to b) till almost vanishing in c).

The particle distribution is inhomogeneous and characterized by
peaks at the walls \cite{maggi2015multidimensional,wagner2017steady,angelani2017confined} whose height is controlled by the persistence time of the force 
orientation, $t_a=1/D_r$ \cite{caprini2019active}.
Indeed, the larger $t_a$, the greater is the time spent by a 
particle in the proximity of the wall and the larger the accumulation 
\cite{lee2013active}.
In the limit of small persistence time ($D_r\gg \gamma$), the accumulation 
becomes negligible and the particles behavior is quite similar to the 
Brownian one, with an effective temperature $T=\gamma U^2_0/2D_r$ 
\cite{fily2012athermal, ten2011brownian, winkler2015virial, kurzthaler2016intermediate}.

To characterize the accumulation degree we report in Fig.\ref{fig:pcond_Y} 
the conditional probability distribution (pdf), $p(y|x)$, 
at two selected vertical sections centered at $x=\pm L/2$. 
The conditional pdf of Fig.\ref{fig:pcond_Y}a, corresponding to 
the snapshot b) in Fig.\ref{fig:Snapshotfree}, displays a bimodal behavior 
with well pronounced peaks due to a marked accumulation of particles at the 
walls. 
The bulk distribution, between the peaks, is not uniform, indicating
that the activity not only promotes the accumulation at the boundaries, 
but it also influences the bulk. 
It is also apparent that the bulk-density is smaller than the density of 
Brownian system counterpart, Fig.\ref{fig:pcond_Y}b. 
This picture is in a qualitative agreement with the prediction of 
Malgaretti and Stark \cite{malgaretti17}. 
When $D_r$  becomes larger enough to determine the approach to the 
Brownian-like regime, 
see Fig.\ref{fig:pcond_Y}a referring to the snapshots 
c) in Fig.\ref{fig:Snapshotfree}, 
the peaks become strongly depleted, and $p(y|x)$ turns to be flat in the 
bulk, as a consequence of a fast transversal homogenization.

The accumulation also depends on the persistence 
length, $\lambda_a = U_0/D_r$, roughly the typical length-scale 
after which particles change direction.  
Indeed, the comparison between $\lambda_a$ and the geometrical sizes 
$H$ and $L$ of the channel, Fig.\ref{fig:wedge}, unveils the interplay between surface 
and bulk properties and allows three main regimes to be identified:
\begin{itemize}

\item[i)] the regime $H \ll L \ll \lambda_a$, where particles move ballistically in all 
directions (Fig.\ref{fig:Snapshotfree}a) and the majority of them 
lay in the proximity of the two walls, eventually sliding along them.\\

\item[ii)] in the regime $H \ll \lambda_a \ll L$, a diffusive 
effective motion emerges along the axis channel while the transversal motion is characterized by rebounds between the walls.
The snapshot b) of Fig.\ref{fig:Snapshotfree} shows that under this condition 
the accumulation reduces and is no longer dominant.

\item[iii)] the regime $\lambda_a \ll H$, where the persistence length is 
smaller than any geometrical scale. 
As shown in Fig.\ref{fig:Snapshotfree}c, the phenomenology is similar to the 
one of a Brownian system at an effective temperature, $T=\gamma U_0^2/2D_r$.
\end{itemize}

As a quantitative measure of accumulation, we plot in Fig.\ref{fig:fig3} the 
fraction $N_w/(N - N_w)$ versus $D_r$, 
by counting the particles contained in the stripes, parallel and adjacent to each boundary, of transversal size $\sigma$.
This ratio exhibits a monotonic decreasing behavior with $D_r$ towards the 
Brownian limit, further
indicating that the increase of $D_r$ depresses the accumulation at the wall. 

\subsection{Case $\epsilon >0$}
We now discuss the case where an external force of strength $\epsilon$ pushes the particles towards the right. 
At variance with the case without drift,  $U_0$ plays a fundamental role 
as it combines with the drift $\epsilon/\gamma$. 
We vary $U_0$, keeping $\epsilon=0.5$,  and explore the two regimes 
$\gamma U_0\geq\epsilon$ and $\gamma U_0 <\epsilon$, at different values of $D_r$. 

In Fig.\ref{fig:Snapshotforce}, we show six snapshots of particle 
configurations at different values of $D_r$ and $U_0$.
Panels a), b) and c) referring to $D_r = 1.0$ and $U_0=1,0.5,0.1$ show a 
Brownian-like behavior with the effective temperature, $T=\gamma U_0^2/2D_r$. 
In this regime, the Brownian fluctuations are not able to counteract the effect of the bias so that no particle can escape on the left.
 
As shown in figures \ref{fig:Snapshotforce}a and \ref{fig:Snapshotforce}b when 
$\epsilon\lesssim \gamma U_0$, the particles may fill vertically the whole 
sector $x>0$ of the channel, whereas in the opposite regime 
($\epsilon\gtrsim \gamma U_0$), the drift prevails over diffusion creating 
a sort of ``plume'' towards the right exit, as illustrated in Fig.\ref{fig:Snapshotforce}c.

The persistent case $D_r=0.01$ is shown in panels d), e) and f). 
In panel d), the particles can explore the whole channel despite the bias, 
on the contrary, when the ratio $\gamma U_0/\epsilon$ decreases, the bias 
prevails, and particles injected at the point $P$ can only explore 
angles $\beta$ such that 
$$
|\beta| \leq \tan^{-1} \bigg(\dfrac{\epsilon}{U_0}\bigg) 
$$
see panels e) and f). 
Such a condition is obtained by assuming a less favorable case 
where the active force has only the $y$-component, thus  
$\dot{x} = \epsilon, \dot{y} = \pm U_0$.

\section{Distribution along channel axis
\label{Sec3}}
\begin{figure}[!t]
\includegraphics[clip=true,width=0.9\columnwidth,keepaspectratio]
{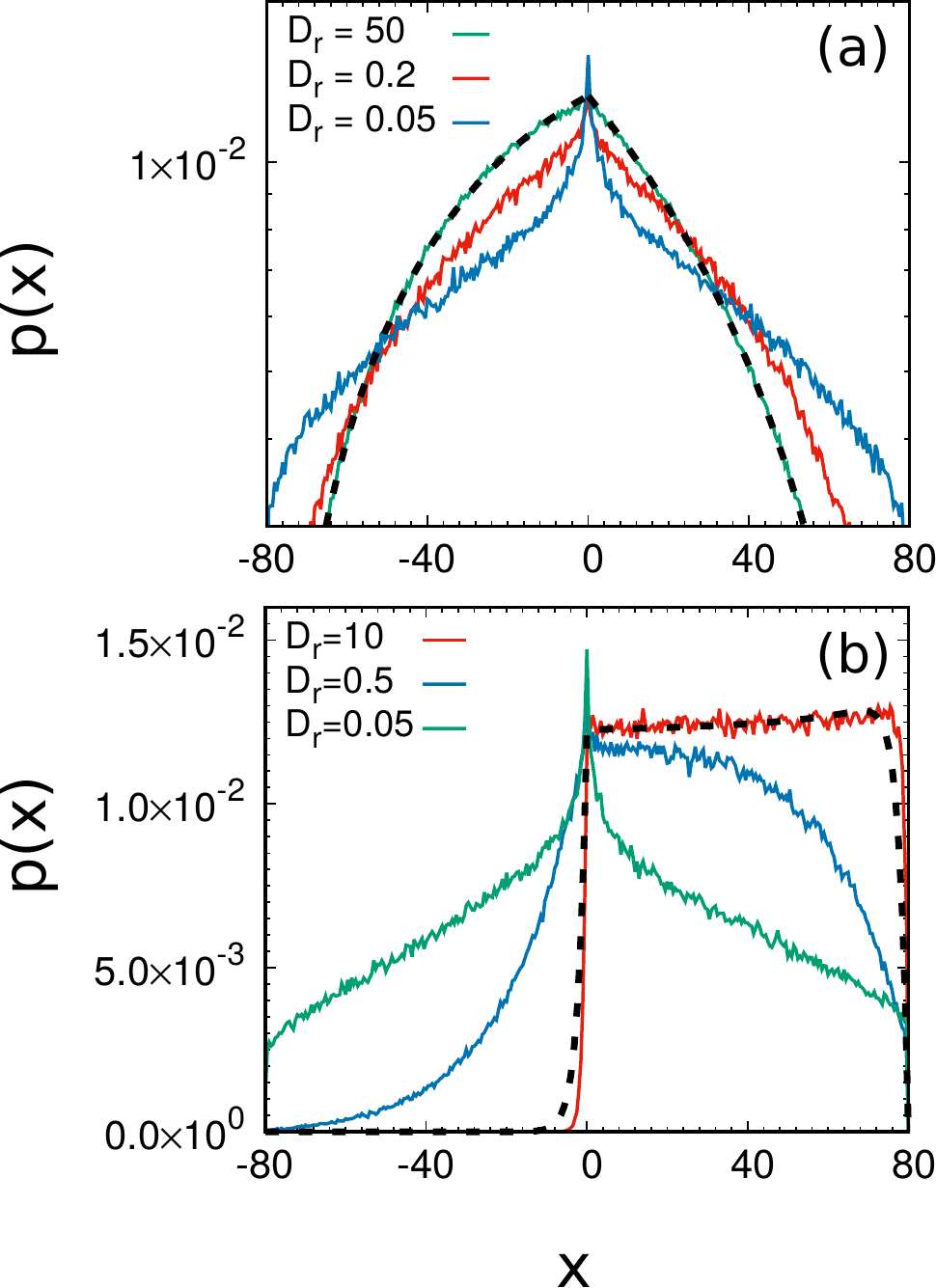}
\caption{Stationary marginal space probability distribution, 
$p_{\mathrm{st}}(x)$, at different values of $D_r$. 
Panel a) refers to $\epsilon=0$, panel b) to $\epsilon>0$; 
dashed lines represent the Brownian predictions: 
Eq.\eqref{eq:statP0} for $\epsilon=0$ and 
Eq.\eqref{eq:eps_statP0} for $\epsilon>0$.
System parameters are: $L=80$, $a=100$, $H=10$, $U_0=3.0$.
\label{fig:PX}}
\end{figure}

A successful approximation often employed in the study of the transport of 
passive particles in narrow channels with non-uniform section 
is represented by the so called Fick-Jacobs approach 
\cite{jacobsBook,zwanzig,burada_Rev}. 
It amounts to reducing the multidimensional 
process to a one-dimensional diffusion in the effective potential encoding 
the channel geometry.
Such an approximation is valid whenever the system reaches a steady
distribution in the transversal section on a time-scale much shorter than 
the typical time of the process along the channel axis 
\cite{buradaTEST,fortePRE,regueraFJ}. 
To what extent this homogenization approach is valid for active particles 
is not clear. Our results show that transversal homogenization is not fulfilled when $D_r < \gamma$, because persistent values of the active force 
favor the accumulation at the walls.
In this respect, we numerically study the stationary marginal probability distribution, 
$p_{\mathrm{st}}(x) $, along the channel axis,
\begin{equation}
p_{\mathrm{st}}(x) = \dfrac{1}{w(x)}\int_{0}^{w(x)}\!\!\!\!dy P_{\mathrm{st}}(x,y) 
\label{eq:Pmarg}
\end{equation}
obtained from the two-dimensional distribution $P_{\mathrm{st}}(x,y)$.
Notice that the existence of a stationary state is a consequence of the  
re-injection that replaces the particles exiting from the boundaries 
$x=\pm L$.
The numerical $p_{\mathrm{st}}(x)$ are reported in Fig.\ref{fig:PX}a in the 
absence of bias and in Fig.\ref{fig:PX}b in the presence of bias, $\epsilon$,
for different values of $D_r$.

For $\epsilon=0$, panel a), we observe an asymmetry with respect to the center of 
the channel ($x=0$) reflecting the narrowing of the section $w(x)$. 
Indeed, the slant of the upper wall generates an ``entropic'' drift 
favoring a larger occupation of the side $x<0$. 
This entropic effect is more evident for large $D_r$ and maximal in the 
Brownian limit characterized, up to a normalization constant, 
by a distribution (dashed black line)
\begin{equation}
p_{st}(x) = 
\begin{cases}
        A  (a - x)\;\ln\bigg(\dfrac{a + L}{a
        -x}\bigg)\qquad & x\in[-L,0] \\
        B  (a - x)\;\ln\bigg(\dfrac{a-x}{a-L}
        \bigg)\qquad & ~~~x\in[0,L]
\end{cases}
\label{eq:statP0}
\end{equation}
predicted by a Fick-Jacobs approach \cite{jacobsBook,burada_Rev,regueraFJ}
that is discussed in detail in appendix \ref{SecBrown}. 
$A, B$ are two coefficients depending on the geometry parameters, $a,L,H$.

It is interesting to remark that formula \eqref{eq:statP0} remains 
reasonably applicable to active particles till to values around $D_r=1$. 
At smaller $D_r$, the entropic drift is contrasted
by the persistence of the trajectories and consequently 
the distribution becomes more symmetric, it also develops a narrow peak near 
$x=0$, more and more pronounced as $D_r$ is reduced.
This over-crowding of the region near $x=0$, absent in the Brownian case, 
is a combined effect of re-injection and persistence that determines the 
accumulation of the particles pointing towards the walls. 

As shown in Fig.\ref{fig:PX}b, a constant field $\epsilon=0.5$ overwhelms 
the ``entropic'' drift and determines a larger density in the region $x>0$. 
Even in this case, the shape of $p_{\mathrm{st}}(x)$ 
is strongly influenced by the activity, 
and again the large $D_r$ range recovers the Brownian-like profile
\begin{equation}
	p_{\mathrm{st}}(x) \propto 
	e^{\varepsilon x} (a-x)
\begin{cases}
	A (\mbox{Ei}[\varepsilon(a+L)] - \mbox{Ei}[\varepsilon(a-x)])\\
	\qquad \qquad \qquad x\in[-L,0] \\
	B (\mbox{Ei}[\varepsilon(a-x)] - \mbox{Ei}[\varepsilon(a-L)])\\
	\qquad \qquad \qquad x\in[0,L]
\end{cases}
\label{eq:eps_statP0}
\end{equation}
which is also derived in Appendix \ref{SecBrown}.
In expression \eqref{eq:eps_statP0}, we set $\varepsilon = \gamma\epsilon/T$ 
(with $T=\gamma U_0^2/2D_r$ meant as an effective temperature) 
and Ei[...] denotes the Exponential Integral function 
(cfr. pp. 661--662 of Ref.\cite{arfken2001}).

We conclude by remarking that in the strong activity regime 
[blue and green curves in Figs. \ref{fig:PX}(a) and \ref{fig:PX}(b)], 
the active force is able to shadow both the entropic and the bias effects, 
thus leading to a symmetrization of the profiles.  

In Sec.\ref{SecEscape}, we see how the accumulation mechanism of the particles 
to the walls strongly affects the escape process.

\section{Escape process of active particles from the wedge}
\label{SecEscape}
We study numerically the escape statistics from the wedge, $Q$, for the ensemble
of particles initially injected in the neighborhood of $(0,H/2)$. 
We define the left and right first passage times, $\tau_{L,R}$, 
as the first time at which a given particle leaves $Q$ either from the 
left or from the right boundary, 
\begin{flalign}
\tau_L&= \min_t{\{0<t\leq T_w \;|\; x(t) <-L}\} \\
\tau_R&= \min_t{\{0<t\leq T_w \;|\; x(t) >L}\}. 
\end{flalign}
within a given simulation time window $[0,T_w]$. 
A convenient choice is $T_{w} \sim 10^4/D_r$ to allow all the particles to exit in a 
reasonable simulation time. 
 We investigate the three different dynamical regimes discussed in Sec.\ref{SecModel}
 and obtain numerically the exit-time distributions, $\Psi_{L,R}(\tau)$, by the histogram  method. 
We start discussing the results in the case of no drift, $\epsilon = 0$, 
and then we consider the driven system, $\epsilon> 0$, using the Brownian 
case as a reference.

\subsection{Active escaping time at $\epsilon=0$}
\begin{figure}[t!]
\centering
\includegraphics[clip=true,width=0.9\columnwidth,keepaspectratio]
{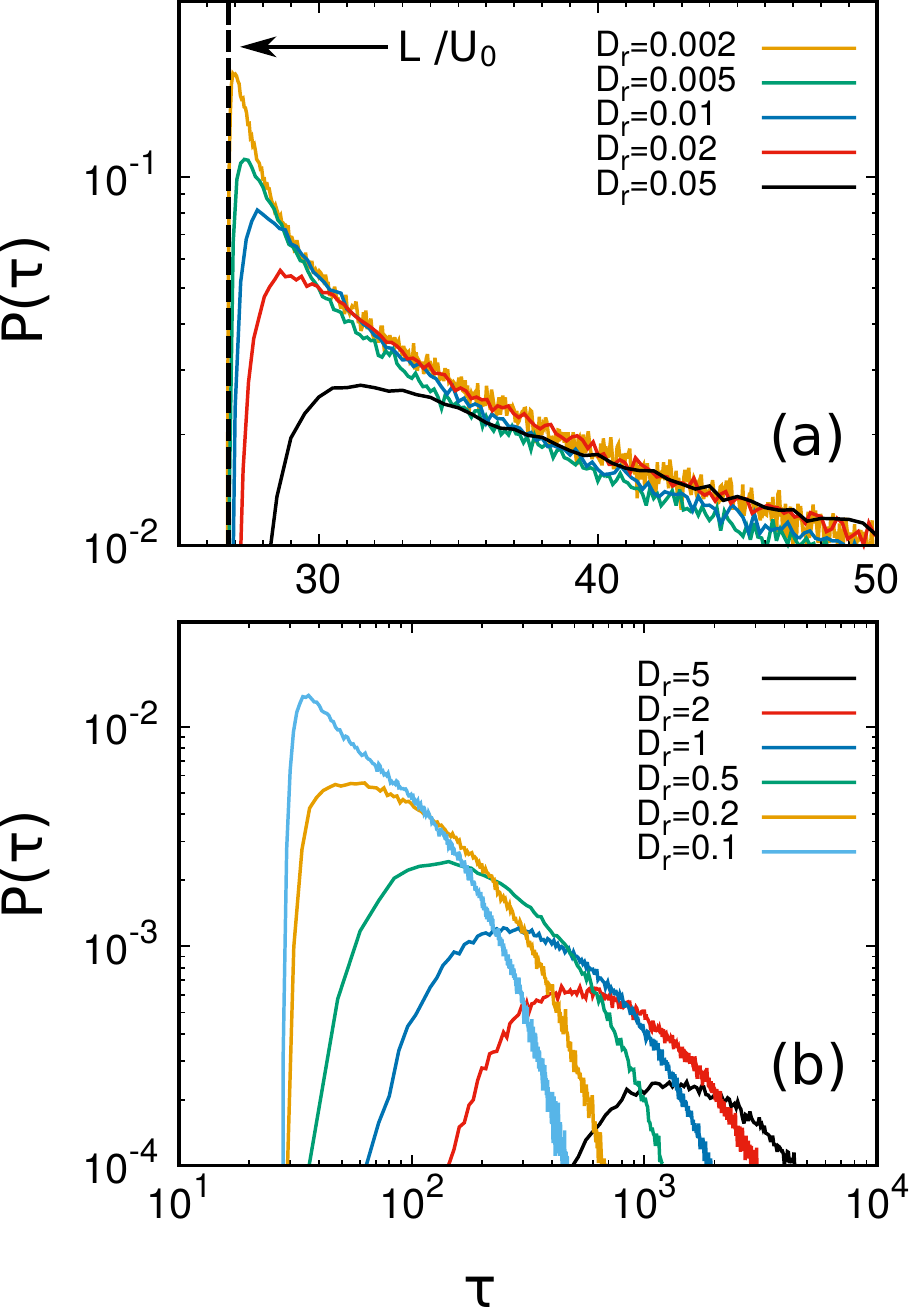}
\caption{Escape time distribution, $\Psi(\tau)$, for $\epsilon = 0$, computed at selected values of $D_r$ in the range $[2 \times 10^{−3} , 5]$, see the legend.
Plots have been split in two panels for readability reasons: panel (a) refers to the range $[2\times 10^{-3}, 5\times 10^{-2}]$ while panel (b) to the range $[0.1, 5.0]$
Other used parameters are $L=80$, $a=100$, $H=10$, and $U_0=3$.
\label{fig:timesfree}
}
\end{figure}

Figure~\ref{fig:timesfree} reports the distributions of first exit times, 
$\Psi(\tau)$, at different values of $D_r$. 

When $\lambda_a \ll H \ll L$, the majority of the particles spreads in the 
bulk and  their behavior is hardly distinguishable from the one of a swarm of  Brownian particles 
with temperature $\gamma \,U_0^2/2D_r$. Accordingly, the $\Psi(\tau)$ is very similar
to the escape time distribution of Brownian particles from the wedge.

When the persistence length is such that, $H\ll\lambda_a \ll 2L$, 
the situation changes because:
i) a large fraction of particles spends much time stuck to the up and bottom 
boundary,  
ii) the vertical component of the particle velocity behaves 
quite ``deterministically'' producing a bouncing ball effects 
between the upper and lower boundaries that lasts for a period of the 
order of the persistence time \cite{lee2013active}, $t_a \sim 1/D_r$.
 As a result of the "stickiness" of the walls, we observe a sort of dimensional
reduction which confers to the $\Psi(\tau)$ a shape strongly deviating 
from the corresponding Brownian distribution.
At first, as  seen in both figures \ref{fig:timesfree}a and 
\ref{fig:timesfree}b, a pronounced asymmetry of $\Psi(\tau)$ occurs, 
characterized by the emergence of a fat tail at larger times and a rather 
steep shoulder at shorter times. 

In the regime $H\ll L \ll \lambda_a$, particles strongly accumulate 
along the walls which act as trails guiding the particles to the left or 
right exit, in a time roughly given by $t_d\sim L /U_0 $. 
As a consequence, the escape problem reduces to the combination of two 
one-dimensional escape processes, each occurring along one of the walls. 
Accordingly, the exit-time distribution becomes extremely peaked near
$t_d$.
The possibility of escaping within $t_d$ clearly depends on the initial 
random orientation of the active force at the injection point. 
In this respect, we can classify the particles in two groups: group A is 
formed by particle with initial direction allowing them to leave the channel 
in a time, $t \sim t_d$, either to the left or to the right, without 
changing direction. 
Instead, group B contains particles changing direction at least one 
time before they reach one of the exits at a larger time. 
The particles belonging to A, arriving quite at the same time $t_d$, 
contribute to the peak in the $\Psi(\tau)$, while the arrivals of the particles of group B contribute to the long tails.
In this regime, the decreasing of $D_r$ produces higher and thinner spikes and longer tails.
\begin{figure}[!t]
\centering
\includegraphics[clip=true,width=0.9\columnwidth,keepaspectratio]
{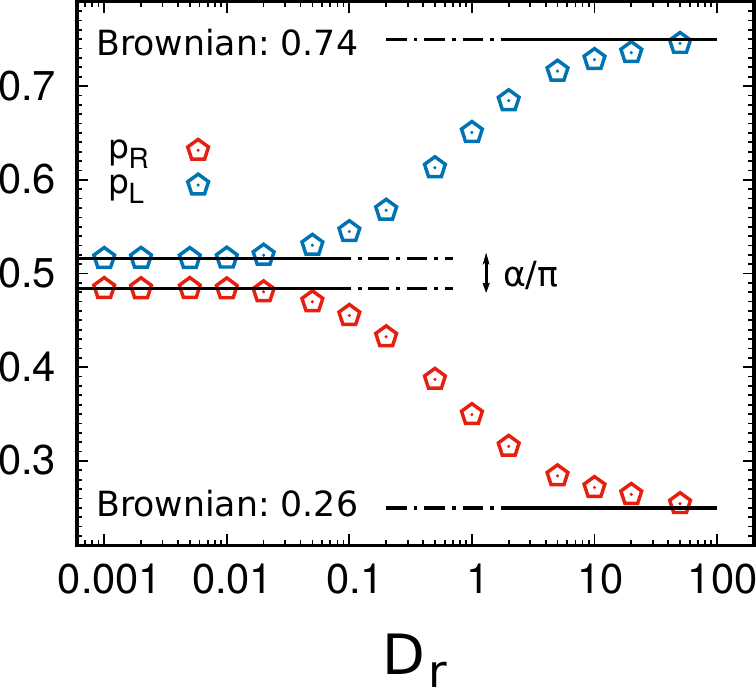}
\caption{Dependence of the left/right escape probabilities, $p_L$ and $p_R$, 
	on $D_r$. 
	Black lines indicate the two limiting plateaus: 
	the Brownian prediction from Eq.\eqref{eq:PBrown}, 
	and the infinite-$\tau$ prediction given by Eq.\eqref{eq:pLR}.
	\label{fig:rate1}}
	\end{figure}
\begin{figure}[!t]
\centering
\includegraphics[clip=true,width=0.9\columnwidth,keepaspectratio]
{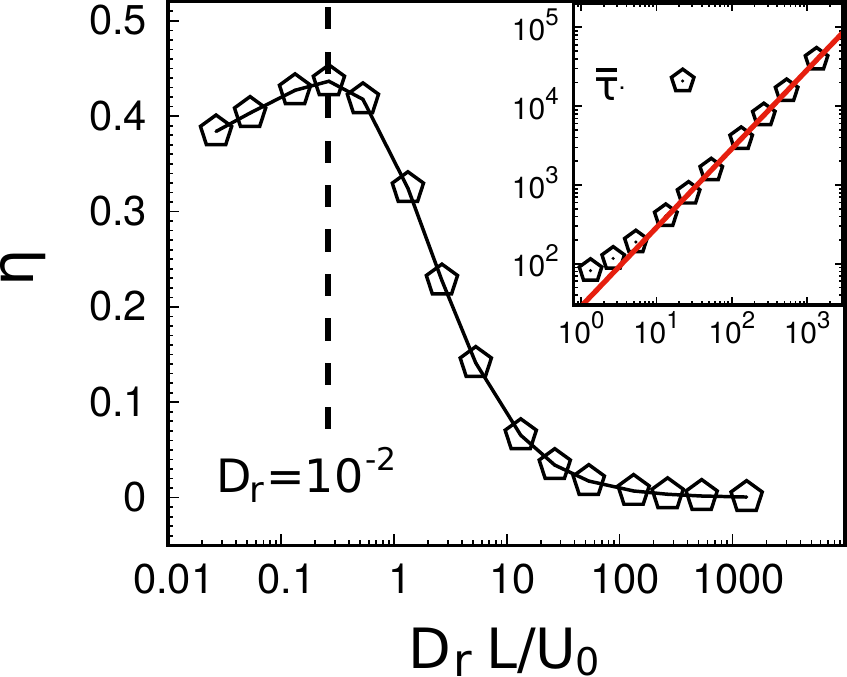}
	\caption{Efficiency, defined by Eq.\eqref{eq:efficiency},
	as a function of the dimensionless parameter $D_r L/U_0$, in the 
	case of zero drift $\epsilon=0$.  
	The peak is attained at $D_r \approx 10^{-2}$. 
	The inset shows $\bar{\tau}$ (black data) vs. $D_r L/U_0$ for a 
	comparison with the Brownian result, Eq.\eqref{eq:taumed}. 
        Parameters: $L=80$, $a=100$, $H=10$, $U_0=3$.
	\label{fig:rate2}}
\end{figure}

\begin{figure*}[!t]
\centering
\includegraphics[clip=true,width=0.8\textwidth,keepaspectratio]
{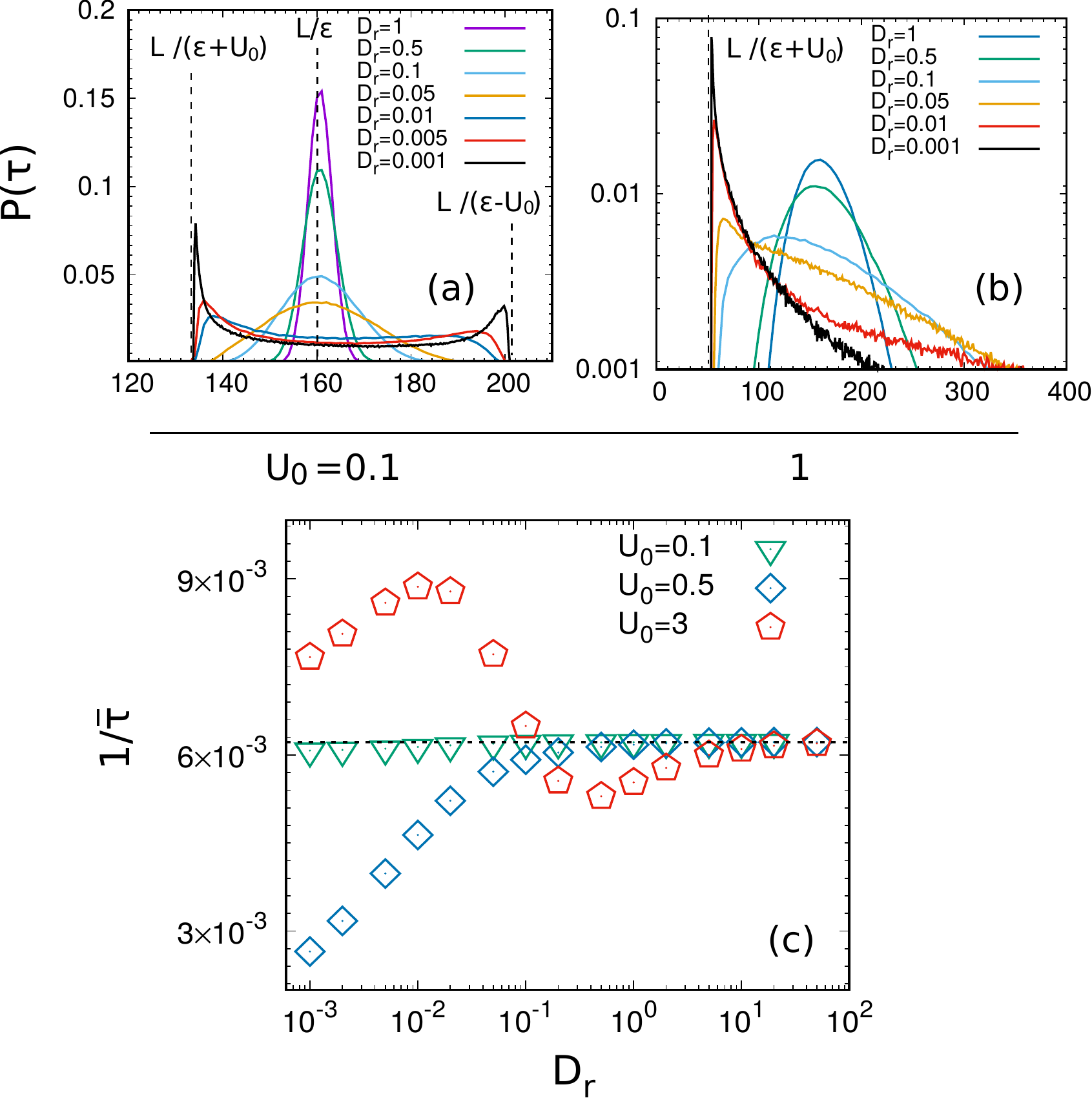}
\caption{Panels a) and b): escape time distributions at different 
values of $D_r$ and two different values of $U_0=0.1,1$ 
in the presence of an external bias $\epsilon=0.5$. 
Panel (c): dependence of $1/\bar{\tau}$, namely the transport efficiency,  
on $D_r$ at values $U_0=0.1,0.5,3$ and $\epsilon=0.5$. 
Geometry parameters are $L=80$, $a=100$, $H=10$.}
\label{fig:Eps-escape}
\end{figure*}

We plot in Fig.\ref{fig:rate1} 
the right and left escaping probability, 
$p_R$, $p_L=1-p_R$, respectively, obtained by measuring the fraction of 
exit events from the right and from the left in a long simulation.
We see that $p_R$ and $p_L$ show a clean monotonic behavior  
as a function of $D_r$, converging to two different plateaus for 
$D_r \rightarrow 0$ and $D_r\rightarrow\infty$, respectively.
The plateau for $D_r \gg \gamma$ (Brownian regime) is given by 
the expression
\begin{equation}
p_R = \dfrac{\ln(a+L) -\ln a}{\ln(a+L) -\ln(a-L)} = \dfrac{\ln(1+ \mu\tan \alpha)}
{\ln\left(\dfrac{1- \mu\tan \alpha}{1 + \mu \tan \alpha}\right)}
\label{eq:PBrown}
\end{equation}
where $\mu = L/H$ and $\alpha$ is the wedge angle, see Fig.\ref{fig:wedge}. 
The derivation of the above expression can be found in 
appendix \ref{SecBrown}, specifically see Eq.\eqref{eq:ratio}.

Notice that in the Brownian case, there is not a temperature dependence.
For $H=100$ and $L=80$, Eq.\eqref{eq:PBrown} provides the values 
$p_R \simeq 0.2675$ which agrees with the simulation value.
In this case, $p_L$ is much less than $p_R$, due to the obvious 
action of the entropic ``drift'' produced by the wedge geometry which favors 
the exit to the larger left side. 
This scenario remains valid up to values of $D_r\sim 10$.

If $D_r$ further decreases, $p_L$ develops a strong dependence on $D_r$,  
indicating that the activity counteracts the entropic drift and facilitates 
the passage through the narrow side of the channel. 
In practice, the particle accumulation to the walls has the effect of 
reducing the entropic barrier. A similar ``rectifying'' 
phenomenology has been observed for active Janus particles in periodic 
channels alternating two wedge compartments \cite{Janus_Ratchet}.

At some value of $D_r$, $p_{R(L)}$ saturates to a value, $p_{R,L}^*$ 
\begin{flalign}
p_{L,R}^* = \frac{\pi \pm \alpha}{2\pi}
\label{eq:pLR}
\end{flalign}
Indeed, if the motion of the particles is so persistent to be 
considered ``ballistic'', $p_L$ and $p_R$ strongly depend on the 
initial re-injection condition. 
With reference to Fig.\ref{fig:wedge}, we can identify two complementary  
intervals $\mathcal{A}_R = [-\pi/2,\pi/2-\alpha]$ and 
$\mathcal{A}_L = [\pi/2-\alpha,3\pi/2]$,    
for which those active particles emitted with an initial angle 
either $\theta_i \in \mathcal{A}_R$ or $\theta_i \in \mathcal{A}_L$ 
are bound to exit almost surely either to the left or to the right, respectively.
Eq.\eqref{eq:pLR} is the analogue of Eq.\eqref{eq:PBrown} in the regime of strong 
persistence. 
Of course, always exists a very small fraction of particles  whose 
orientation prevents the exit in a time $t_a$, but this fraction is very small 
and does not affect the exit time statistics except for the short tail.

It is interesting to study the average exit time from the wedge 
$$
\bar{\tau}= \int_0^{\infty}\!\!\!d\tau \Psi(\tau) \tau
$$
as a function of the control parameters.
This observable is able to quantify the transport efficiency and it is 
relevant to understand if the activity favors or not the emptying of the 
channel. 

We can define the transport efficiency as the ratio 
\begin{equation}
\eta = \dfrac{L}{U_0\bar{\tau}} 
	\label{eq:efficiency}
\end{equation}
between the time, $L/U_0$, at which a deterministic motion of velocity 
$U_0$  gains the exit and the mean exit time $\bar{\tau}$.

In Fig.\ref{fig:rate2}, we plot $\eta$ for $U_0=3$ 
versus the dimensionless parameter $D_r L/U_0$. 
It exhibits a non-monotonous behavior reaching its maximal value at 
$D_r L/U_0\approx 0.3$.
This reveals the existence of an optimal $D_r$ such that the escaping process 
becomes more efficient, in the specific parameter choice $D_r \simeq 0.01$. 
The increase of the ``transport efficiency'' can be explained by invoking a 
sort of ``dimensional reduction''. 
Specifically, those particles accumulating at the walls 
are favored in the exit process because they use the boundaries like trails, 
thus performing basically a one-dimensional motion along them. 
This greatly enhances the possibility to find the exit with respect to the case where particles explore the full wedge in order to escape.
Ref.~\cite{malgaretti17} shows that the accumulation near the 
walls also depends on the hydrodynamics interactions.
As a consequence, the efficiency peak in Fig.9 shifts towards larger  
or smaller values of $D_r L/U_0$ in the case of pullers or pushers, 
respectively.

At very small $D_r L/U_0$, however $1/\bar{\tau}$ decreases as a finite fraction of 
particles, in particular, those hitting normally the walls, almost remain stuck for a time $1/D_r$,
(diverging for $D_r \to 0$) thus slowing down their escape process.
Again for $D_r \gtrsim \gamma$, the system approaches a 
Brownian regime and $\bar{\tau}$ linearly increases with $D_r L/U_0$, as shown in the inset of Fig.\ref{fig:rate2}. 
Indeed, the effective temperature  $T=\gamma U_0^2/2D_r$, which in this case 
controls the Brownian-like behavior, increases with $D_r L/U_0$. 
The inset of Fig.\ref{fig:rate2} shows also a comparison between the numerical 
$\bar{\tau}$ and the prediction derived in appendix \ref{SecBrown}
for a  Brownian particle at temperature $T=\gamma U_0^2/2D_r$:
\begin{equation}
	\bar{\tau} = \frac{D_r L^2}{2\gamma U_0^2}\left(1 + \frac{a}{2 L} 
	\dfrac{\ln[a^2/(a^2 - L^2)]}{\ln[(a + L)/(a - L)]}\right).
	\label{eq:taumed}
\end{equation}
The first term corresponds to the average exit time of a one-dimensional system  
of length $2L$, whereas the second one is associated with the entropic barrier and reflects the asymmetry of the channel.
As expected, the linear behavior of  $\bar{\tau}$ with $D_r$ in the 
effective equilibrium regime is in very good agreement  with the prediction.

\subsection{Active escaping time at $\epsilon>0$}
In this section, we analyze how an external driving $\epsilon$ modifies the previous scenario.
Consistently with the concept of effective temperature, when $D_r$ is large enough 
we expect a Brownian-like regime and thus an exit time distribution, $\Psi(\tau)$,  resembling the corresponding Brownian distribution at temperature $\gamma\, U_0^2/2D_r$, as shown in Fig.\ref{fig:Eps-escape}. 
In this case $\Psi(\tau)$ is peaked around, $t_m \sim \gamma L/\epsilon$, 
representing the time taken by particles of velocity $\epsilon/\gamma$ to travel a distance $L$.
In this regime, the reduction of $U_0$ or $D_r$ increases the variance of $\Psi(\tau)$. 
In the Brownian-like regime, the decreasing of $D_r$ produces the enhancement 
of the skewness of $\Psi(\tau)$ and the emergence of long right tails.

A further decrease of $D_r$ shifts the peak of $\Psi(\tau)$ to the left and 
simultaneously the right tail becomes higher, until the mean peak position pins at $t_m \sim \gamma L/(\gamma U_0+\epsilon)$. 
In this regime, the reduction of $D_r$ leads only to more pronounced peaks of $\Psi(\tau)$. 
Indeed, even the particles which move ballistically towards the exit without changing 
their orientation (group A) cannot reach the exit within the minimal time  $\sim \gamma L/(\gamma U_0+\epsilon)$. 
Depending on the ratio $\gamma U_0/\epsilon$ a different phenomenology occurs:
i) $\gamma U_0<\epsilon$: a secondary peak of $\Psi(\tau)$ occurs at a value $\gamma L/(\epsilon- \gamma U_0)$ and finally the distribution 
abruptly drops down.
This secondary peak is due to the slow particles whose orientation is opposed 
to the $x$-direction where the constant force is directed.
ii) $\gamma U_0>\epsilon$: the first peak simply becomes higher but the 
second vanishes.

The efficiency $\eta$ of the transport for $\epsilon > 0$ 
is defined by replacing $U_0\to U_0 + \epsilon$ in Eq.\eqref{eq:efficiency}, 
but as a matter of fact, the inverse of the mean exit time is already 
an estimate of $\eta$. For this reason Fig.\ref{fig:Eps-escape}c reports 
directly $1/\bar{\tau}$ versus $D_r$ to quantify 
the channel emptying at three values of $U_0$. 
In the case $\epsilon > \gamma U_0$, the increasing of $D_r$, 
i.e. the decreasing of the effective temperature, 
leads to a monotonic growth of $1/\bar{\tau}$, until  saturation is reached 
and the system behaves as a Brownian one.
In this regime, the activity reduces the efficiency of the transport process. 
Indeed, despite some particles travel towards the exit with more facility, also several particles move in 
the opposite direction, employing a long time before leaving the wedge.
Clearly, the reduction effect becomes relevant only when $U_0$ is comparable with $\epsilon$ (blue diamond data) 
otherwise (when $\gamma U_0 \ll \epsilon$) the particles leave the 
wedge only because of the driving $\epsilon$ (green triangle data).

The situation is more interesting in the opposite regime, 
where $\epsilon < \gamma U_0$. 
In this case,  $1/\bar{\tau}$, reveals a non-monotonic behavior in terms of $D_r$.  
Starting from the Brownian saturation value, a first decreasing of $D_r$ produces a reduction of 
 $1/\bar{\tau}$, until a minimum value; a situation resembling the previous one for $\epsilon> \gamma U_0$. 
Nevertheless, a further decrease of $D_r$, produces the increasing of $1/\bar{\tau}$, up to a maximum value 
which reveals an increase of the transport efficiency for some value of $D_r$. 
This persistence maximizes the efficiency of the transport process, since for 
smaller $D_r$ a decrease of $1/\bar{\tau}$ occurs, due to the same mechanism 
already discussed in the case $\epsilon=0$. 

As a consequence, the activity can be seen as an optimization mechanism, 
which reduces the time employed by microswimmers to reach the exit of the 
channel, also in the presence of a constant driving force. 
A condition for this scenario implies that the activity strength has to be 
stronger than the amplitude of the external driving.

\section{Conclusions}
\label{SecConclusions}
In this work, we studied the escape process of a system of ABP particles
from an open-wedge channel in the absence and in the presence of an external 
driving force. 
The comparison with the Brownian system counterpart shows that activity 
facilitates the escaping from the narrow exit and competes against the 
entropic force. 
We also found the existence of an optimum value of the persistence time which maximizes the channel emptying.
The physical mechanism which improves the efficiency is the effective 
attraction exerted by the bottom and upper walls on the active particles 
which leads to a depletion of the inner region. 
The majority of the particles accumulate in a narrow region near each wall and as a consequence, the section-dependent entropic 
force is strongly reduced. 
The resulting motion of the particles is effectively one-dimensional 
and controlled by the competition between the active force and the external 
field. 
Somehow, the activity is able to operate a sort of dimensional reduction, since makes surface effects prevailing over the bulk properties.

The present treatment shows that the extension of the Fick-Jacobs approximation to the active case is possible only for small values 
of the persistent time and activity strength \cite{dagdug14}. 
When it happens, the far-equilibrium feature of active particles do not emerge and the Brownian theory with an effective temperature 
agrees with numerical data.
On the contrary, when the active force is very persistent and large, our study of particle distributions shows that the main hypothesis underlying 
the Fick-Jacobs approximation breaks down, since the density is not homogeneous in each channel section, 
as an effect of the accumulation near the walls. 
Quantifying how the effective entropic barrier is reduced by this mechanism could represent an intriguing next point in understanding 
biological transport processes. 

This work suggests that the active transport can be facilitated or 
even optimized by a proper design of the channel surface. 
This can be interpreted as further manifestation of ``ratcheting mechanism'' 
that can be observed as soon as active particles experience confining forces 
breaking the left-right symmetry~\cite{Janus_Ratchet}.

We point out that the presence of inter-particle interactions could drastically 
affect the transport properties in the channel. Indeed, as mentioned in 
the introduction, interactions could promote clustering until 
MIPS takes place in the regime 
of strong active forces. 
Assessing the effects of MIPS on the mean exit time from the channel 
will certainly constitute an interesting subject for future investigations.

\appendix
\section{Brownian case \label{SecBrown}}
In this appendix, we describe the spreading of Brownian particles 
distribution, $P(x,t)$, along the axis of the channel $Q$ in terms 
of Fick-Jacobs approach \cite{jacobsBook}. Fick-Jacobs theory is applicable 
to channels with variable sections, $w(x)$, provided the particle distribution
attains its steady form along the $y$-direction on a time-scale much shorter than the time scale associated with the 
longitudinal motion (transversal homogenization).
However, this condition, that can be fulfilled by Brownian particles,  
is violated by the active particles in those regimes where accumulation to the boundaries 
is not negligible. 

In the Brownian regimes, the strong confinement in the lateral direction 
allows the diffusion of the particles along the channel to be described in a quasi-one-dimensional equation
\begin{align}
& \dfrac{\partial P}{\partial t} +  
\dfrac{\partial J}{\partial x} = k\;\delta(x) \\
     \nonumber  \\
&P(L,t) = P(-L,t) = 0
\end{align}
where the last two equations mathematically implement 
the absorbing boundary conditions at $x=\pm L$.
The Dirac-delta source with amplitude $k$ accounts for 
the instantaneous re-injection, at $x=0$, of those particles 
leaving the channel from the boundaries $x=\pm L$.

The current 
\begin{equation}
	J(x,t) = -\dfrac{T}{\gamma} \,e^{\gamma\epsilon x/T} w(x)
        \dfrac{\partial}{\partial x} \bigg[e^{-\gamma\epsilon x/T}\dfrac{P(x,t)}{w(x)} 
        \bigg]
\label{eq:curreps}
\end{equation}
describes to a good approximation the longitudinal transport  
along a channel of variable section $w(x) = H - H x/a$, 
in the presence of a constant field (bias) of strength $\epsilon$ 
acting along the channel axis. 
In the following, we set 
$$
D_0 = \dfrac{T}{\gamma} \quad, \quad \varepsilon = \dfrac{\gamma\epsilon}{T}\;.
$$
The balance between absorption and re-injection 
preserves the number of particles and gives rise to a
steady-state characterized by the equality 
$$J(L) - J(-L) = k,$$ stating that the re-injection rate balances
the loss fluxes at the boundaries.

 The stationary distribution satisfies the equation 
\begin{equation}
-D_0 \dfrac{\partial}{\partial x} 
\bigg[
w(x)e^{\varepsilon x} \dfrac{\partial}{\partial x} e^{-\varepsilon x} \dfrac{P}{w(x)} 
\bigg] = k\;\delta(x).
\label{eq:static}
\end{equation}
The presence of the Dirac $\delta(x)$ requires the splitting of the solution 
over the two domains $[-L,0]$ and $[0,L]$, such that:
\begin{equation}
P(x) = 
\begin{cases}
  A\; Y_1(x)  \qquad & -L\leq x <0 \\
  B\; Y_2(x)  \qquad & ~~~0\leq x \leq L
\end{cases}
\label{eq:P0}
\end{equation}
where
$$
Y_1(x) = e^{\varepsilon x} w(x) \int_{-L}^x\!du 
\dfrac{e^{-\varepsilon u}}{w(u)} 
$$
and 
$$
Y_2(x) = e^{\varepsilon x} w(x)\int_{x}^L\!du \dfrac{e^{-\varepsilon u}}{w(u)} 
$$
are the fundamental solutions of the homogeneous equation ($k=0$) 
satisfying the boundary conditions $Y_1(-L) = Y_2(L) = 0$.  
The coefficients $A,B$ are determined by imposing the continuity 
condition, $P(0^{+}) = P(0^{-})$, and from integrating both members
of Eq.\eqref{eq:static} over the interval $[-\Delta,\Delta]$, then taking 
$\Delta\to 0$. 

The solution of these constraints provides, 
$$
A = \int_{0}^{L}\!dx\dfrac{e^{-\varepsilon x}}{w(x)} 
\quad,\quad 
B = \int_{-L}^{0}\!dx\dfrac{e^{-\varepsilon x}}{w(x)}.
$$
A little algebraic manipulation yields the explicit result
\begin{eqnarray}
A =\mbox{Ei}[\varepsilon a] - \mbox{Ei}[\varepsilon (a-L)] \\
B =\mbox{Ei}[\varepsilon(a+L)] - \mbox{Ei}[\varepsilon a], 
\end{eqnarray}
where $\mbox{Ei}[u]$ is the Exponential Integral of argument $u$ 
\cite{arfken2001}.
The final expression reads
\begin{equation}
	P(x) \propto e^{\varepsilon x} (a-x)  
\begin{cases}
	A(\mbox{Ei}[\varepsilon(a+L)] - \mbox{Ei}[\varepsilon(a-x)])\\
	 \qquad \mbox{when~} x\in[-L,0] \\
	B(\mbox{Ei}[\varepsilon(a-x)] - \mbox{Ei}[\varepsilon(a-L)]) \\
	\qquad \mbox{when~} x\in[0,L]
\end{cases}
\label{eq:finalP0}
\end{equation}
up to a normalization constant such that the integral of $P(x)$ over $[-L,L]$ 
is set to $1$.

These expressions drastically simplify in the zero-field limit, $\epsilon =0$,
\begin{equation}
P(x) = C(a -x)
\begin{cases}
     A\;\ln\left[\dfrac{a + L}{a-x}\right] & \quad x\in[-L,0] \\
     B\;\ln\left[\dfrac{a - x}{a-L}\right] & \quad x\in[0,L]\\
\end{cases}
\label{eq:eps_finalP0}
\end{equation}
with  
$A = \ln[(a + L)/a]$, $B = \ln[a/(a - L)]$ and a normalization constant
$$
C^{-1} = \dfrac{L}{4} \left[2 a \ln
   \left(\dfrac{a^2}{a^2-L^2}\right)+L \ln
   \left(\dfrac{a+L}{a-L}\right)\right].
$$
Now, we are in the position to derive formula \eqref{eq:PBrown} for the ratio $p_R$, 
which determines the Brownian limit in Fig.\ref{fig:rate1}. 
The value of $p_R$ can be estimated in term of the fluxes at the boundaries, which, 
due to the fact that $P(\pm L) = 0$ are only proportional to the derivatives of $P$ 
at $x=\pm L$, i.e. $J(\pm L) = -D_0 P'(\pm L)$. 
Then $p_R$ is obtained as the fraction of the flux to the right over the total 
flux   
\begin{equation}
p_R = \dfrac{J(L)}{J(L) + J(-L)} = 
\dfrac{\ln(a + L) - \ln(a)}{\ln(a + L) - \ln(a - L)} .
\label{eq:ratio}
\end{equation}\\

The average first-arrival time at the boundaries $\pm L$,  
for a particle released at $x$, is related to the survival probability, 
$S(x,t)$, by the integral \cite{Gardiner}
$$
\bar{\tau}(x) = \int_{0}^{\infty}\!\!dt\;S(x,t)\;. 
$$
By definition $S(x,t)$ is the probability that the particle has not yet left 
the interval $[-L,L]$ at the time $t$ and it is known to satisfy the backward 
Fokker-Planck equation \cite{Gardiner}
\begin{equation}
\dfrac{\partial S}{\partial t} = 
\dfrac{D_0}{w(x)} \dfrac{\partial}{\partial x} 
\bigg[w(x) \dfrac{\partial S}{\partial x} 
       \bigg] 
\label{eq:Survive}
\end{equation}
with the boundary conditions $S(\pm L,t) = 0$.
To obtain a differential equation for $\bar{\tau}(x)$ it is sufficient to integrate 
Eq.\eqref{eq:Survive} in the interval $0\le t < \infty$, and taking into account 
that $S(x,\infty) = 0$ and $S(x,0) = 1$, thus 
\begin{equation}
\dfrac{d}{dx}\bigg[w(x) \dfrac{d\bar{\tau}}{dx}\bigg] = -\dfrac{w(x)}{D_0} ,
\end{equation}
that has to be solved with the obvious boundary values $\bar{\tau}(\pm L) = 0$,
stating that particle emitted at the boundary are instantaneously absorbed.
We are interested in the the average arrival time from $x=0$, then the searched 
solution is
$$
\bar{\tau}(x=0) = \dfrac{L^2}{4D_0} + \dfrac{a L}{2 D_0}\; 
\dfrac{\ln[a^2/(a^2 - L^2)]}{\ln[(a + L)/(a - L)]}
$$
which is exactly Eq.\eqref{eq:taumed}.


\begin{thebibliography}{0}%
\makeatletter
\providecommand \@ifxundefined [1]{%
 \@ifx{#1\undefined}
}%
\providecommand \@ifnum [1]{%
 \ifnum #1\expandafter \@firstoftwo
 \else \expandafter \@secondoftwo
 \fi
}%
\providecommand \@ifx [1]{%
 \ifx #1\expandafter \@firstoftwo
 \else \expandafter \@secondoftwo
 \fi
}%
\providecommand \natexlab [1]{#1}%
\providecommand \enquote  [1]{``#1''}%
\providecommand \bibnamefont  [1]{#1}%
\providecommand \bibfnamefont [1]{#1}%
\providecommand \citenamefont [1]{#1}%
\providecommand \href@noop [0]{\@secondoftwo}%
\providecommand \href [0]{\begingroup \@sanitize@url \@href}%
\providecommand \@href[1]{\@@startlink{#1}\@@href}%
\providecommand \@@href[1]{\endgroup#1\@@endlink}%
\providecommand \@sanitize@url [0]{\catcode `\\12\catcode `\$12\catcode
  `\&12\catcode `\#12\catcode `\^12\catcode `\_12\catcode `\%12\relax}%
\providecommand \@@startlink[1]{}%
\providecommand \@@endlink[0]{}%
\providecommand \url  [0]{\begingroup\@sanitize@url \@url }%
\providecommand \@url [1]{\endgroup\@href {#1}{\urlprefix }}%
\providecommand \urlprefix  [0]{URL }%
\providecommand \Eprint [0]{\href }%
\providecommand \doibase [0]{http://dx.doi.org/}%
\providecommand \selectlanguage [0]{\@gobble}%
\providecommand \bibinfo  [0]{\@secondoftwo}%
\providecommand \bibfield  [0]{\@secondoftwo}%
\providecommand \translation [1]{[#1]}%
\providecommand \BibitemOpen [0]{}%
\providecommand \bibitemStop [0]{}%
\providecommand \bibitemNoStop [0]{.\EOS\space}%
\providecommand \EOS [0]{\spacefactor3000\relax}%
\providecommand \BibitemShut  [1]{\csname bibitem#1\endcsname}%
\let\auto@bib@innerbib\@empty
\end{thebibliography}%


\begin{mcitethebibliography}{67}
\providecommand*{\natexlab}[1]{#1}
\providecommand*{\mciteSetBstSublistMode}[1]{}
\providecommand*{\mciteSetBstMaxWidthForm}[2]{}
\providecommand*{\mciteBstWouldAddEndPuncttrue}
 {\def\EndOfBibitem{\unskip.}}
\providecommand*{\mciteBstWouldAddEndPunctfalse}
  {\let\EndOfBibitem\relax}
\providecommand*{\mciteSetBstMidEndSepPunct}[3]{}
\providecommand*{\mciteSetBstSublistLabelBeginEnd}[3]{}
\providecommand*{\EndOfBibitem}{}
\mciteSetBstSublistMode{f}
\mciteSetBstMaxWidthForm{subitem}
{(\emph{\alph{mcitesubitemcount}})}
\mciteSetBstSublistLabelBeginEnd{\mcitemaxwidthsubitemform\space}
{\relax}{\relax}

\bibitem[Berg(2008)]{berg2008coli}
H.~Berg, \emph{E. Coli in Motion}, Springer Science \& Business Media,
  2008\relax
\mciteBstWouldAddEndPuncttrue
\mciteSetBstMidEndSepPunct{\mcitedefaultmidpunct}
{\mcitedefaultendpunct}{\mcitedefaultseppunct}\relax
\EndOfBibitem
\bibitem[Blake and Sleigh(1974)]{blake1974mechanics}
J.~R. Blake and M.~A. Sleigh, \emph{Biol. Rev.}, 1974, \textbf{49},
  85--125\relax
\mciteBstWouldAddEndPuncttrue
\mciteSetBstMidEndSepPunct{\mcitedefaultmidpunct}
{\mcitedefaultendpunct}{\mcitedefaultseppunct}\relax
\EndOfBibitem
\bibitem[Woolley(2003)]{woolley2003motility}
D.~Woolley, \emph{Reproduction}, 2003, \textbf{126}, 259--270\relax
\mciteBstWouldAddEndPuncttrue
\mciteSetBstMidEndSepPunct{\mcitedefaultmidpunct}
{\mcitedefaultendpunct}{\mcitedefaultseppunct}\relax
\EndOfBibitem
\bibitem[Poujade \emph{et~al.}(2007)Poujade, Grasland-Mongrain, Hertzog,
  Jouanneau, Chavrier, Ladoux, Buguin, and Silberzan]{poujade2007collective}
M.~Poujade, E.~Grasland-Mongrain, A.~Hertzog, J.~Jouanneau, P.~Chavrier,
  B.~Ladoux, A.~Buguin and P.~Silberzan, \emph{Proc. Natl. Acad. Sci. USA},
  2007, \textbf{104}, 15988--15993\relax
\mciteBstWouldAddEndPuncttrue
\mciteSetBstMidEndSepPunct{\mcitedefaultmidpunct}
{\mcitedefaultendpunct}{\mcitedefaultseppunct}\relax
\EndOfBibitem
\bibitem[K{\"o}hler \emph{et~al.}(2011)K{\"o}hler, Schaller, and
  Bausch]{kohler2011structure}
S.~K{\"o}hler, V.~Schaller and A.~R. Bausch, \emph{Nature materials}, 2011,
  \textbf{10}, 462\relax
\mciteBstWouldAddEndPuncttrue
\mciteSetBstMidEndSepPunct{\mcitedefaultmidpunct}
{\mcitedefaultendpunct}{\mcitedefaultseppunct}\relax
\EndOfBibitem
\bibitem[Walther and Müller(2013)]{walther2013janus}
A.~Walther and A.~H. Müller, \emph{Chem. Rev.}, 2013, \textbf{113},
  5194--5261\relax
\mciteBstWouldAddEndPuncttrue
\mciteSetBstMidEndSepPunct{\mcitedefaultmidpunct}
{\mcitedefaultendpunct}{\mcitedefaultseppunct}\relax
\EndOfBibitem
\bibitem[Lattuada and Hatton(2011)]{lattuada2011synthesis}
M.~Lattuada and T.~Hatton, \emph{Nano Today}, 2011, \textbf{6}, 286--308\relax
\mciteBstWouldAddEndPuncttrue
\mciteSetBstMidEndSepPunct{\mcitedefaultmidpunct}
{\mcitedefaultendpunct}{\mcitedefaultseppunct}\relax
\EndOfBibitem
\bibitem[Bechinger \emph{et~al.}(2016)Bechinger, Di~Leonardo, Lowen,
  Reichhardt, and Volpe]{bechinger2016active}
C.~Bechinger, R.~Di~Leonardo, H.~Lowen, C.~Reichhardt and G.~Volpe, \emph{Rev.
  Mod. Phys.}, 2016,  045006(50)\relax
\mciteBstWouldAddEndPuncttrue
\mciteSetBstMidEndSepPunct{\mcitedefaultmidpunct}
{\mcitedefaultendpunct}{\mcitedefaultseppunct}\relax
\EndOfBibitem
\bibitem[Romanczuk \emph{et~al.}(2012)Romanczuk, B{\"a}r, Ebeling, Lindner, and
  Schimansky-Geier]{romanczuk2012active}
P.~Romanczuk, M.~B{\"a}r, W.~Ebeling, B.~Lindner and L.~Schimansky-Geier,
  \emph{Eur. Phys. J. Special Topics}, 2012, \textbf{202}, 1--162\relax
\mciteBstWouldAddEndPuncttrue
\mciteSetBstMidEndSepPunct{\mcitedefaultmidpunct}
{\mcitedefaultendpunct}{\mcitedefaultseppunct}\relax
\EndOfBibitem
\bibitem[Marchetti \emph{et~al.}(2013)Marchetti, Joanny, Ramaswamy, Liverpool,
  Prost, Rao, and Simha]{marchetti2013hydrodynamics}
M.~Marchetti, J.~Joanny, S.~Ramaswamy, T.~Liverpool, J.~Prost, M.~Rao and R.~A.
  Simha, \emph{Rev. Mod. Phys.}, 2013, \textbf{85}, 1143--1189\relax
\mciteBstWouldAddEndPuncttrue
\mciteSetBstMidEndSepPunct{\mcitedefaultmidpunct}
{\mcitedefaultendpunct}{\mcitedefaultseppunct}\relax
\EndOfBibitem
\bibitem[Ramaswamy(2010)]{ramaswamy2010mechanics}
S.~Ramaswamy, \emph{Annu. Rev. Condens. Matter Phys.}, 2010, \textbf{1},
  323--345\relax
\mciteBstWouldAddEndPuncttrue
\mciteSetBstMidEndSepPunct{\mcitedefaultmidpunct}
{\mcitedefaultendpunct}{\mcitedefaultseppunct}\relax
\EndOfBibitem
\bibitem[Nash \emph{et~al.}(2010)Nash, Adhikari, Tailleur, and
  Cates]{nash2010run}
R.~Nash, R.~Adhikari, J.~Tailleur and M.~Cates, \emph{Phys. Rev. Lett.}, 2010,
  \textbf{104}, 258101\relax
\mciteBstWouldAddEndPuncttrue
\mciteSetBstMidEndSepPunct{\mcitedefaultmidpunct}
{\mcitedefaultendpunct}{\mcitedefaultseppunct}\relax
\EndOfBibitem
\bibitem[Tailleur and Cates(2008)]{tailleur2008statistical}
J.~Tailleur and M.~Cates, \emph{Phys. Rev. Lett.}, 2008, \textbf{100},
  218103\relax
\mciteBstWouldAddEndPuncttrue
\mciteSetBstMidEndSepPunct{\mcitedefaultmidpunct}
{\mcitedefaultendpunct}{\mcitedefaultseppunct}\relax
\EndOfBibitem
\bibitem[Sevilla and Nava(2014)]{sevilla2014theory}
F.~J. Sevilla and L.~A.~G. Nava, \emph{Phys. Rev. E}, 2014, \textbf{90},
  022130\relax
\mciteBstWouldAddEndPuncttrue
\mciteSetBstMidEndSepPunct{\mcitedefaultmidpunct}
{\mcitedefaultendpunct}{\mcitedefaultseppunct}\relax
\EndOfBibitem
\bibitem[ten Hagen \emph{et~al.}(2011)ten Hagen, van Teeffelen, and
  L{\"o}wen]{ten2011brownian}
B.~ten Hagen, S.~van Teeffelen and H.~L{\"o}wen, \emph{J. Phys. Condens.
  Matter}, 2011, \textbf{23}, 194119\relax
\mciteBstWouldAddEndPuncttrue
\mciteSetBstMidEndSepPunct{\mcitedefaultmidpunct}
{\mcitedefaultendpunct}{\mcitedefaultseppunct}\relax
\EndOfBibitem
\bibitem[Szamel(2014)]{szamel2014self}
G.~Szamel, \emph{Phys. Rev. E}, 2014, \textbf{90}, 012111\relax
\mciteBstWouldAddEndPuncttrue
\mciteSetBstMidEndSepPunct{\mcitedefaultmidpunct}
{\mcitedefaultendpunct}{\mcitedefaultseppunct}\relax
\EndOfBibitem
\bibitem[Das \emph{et~al.}(2018)Das, Gompper, and Winkler]{das2018confined}
S.~Das, G.~Gompper and R.~Winkler, \emph{New J. Phys.}, 2018, \textbf{20},
  015001\relax
\mciteBstWouldAddEndPuncttrue
\mciteSetBstMidEndSepPunct{\mcitedefaultmidpunct}
{\mcitedefaultendpunct}{\mcitedefaultseppunct}\relax
\EndOfBibitem
\bibitem[Marconi Marini~Bettolo and Maggi(2015)]{marconi2015towards}
U.~Marconi Marini~Bettolo and C.~Maggi, \emph{Soft Matter}, 2015, \textbf{11},
  8768--8781\relax
\mciteBstWouldAddEndPuncttrue
\mciteSetBstMidEndSepPunct{\mcitedefaultmidpunct}
{\mcitedefaultendpunct}{\mcitedefaultseppunct}\relax
\EndOfBibitem
\bibitem[Mi\~{n}o \emph{et~al.}(2018)Mi\~{n}o, Baabour, Chertcoff, Gutkind,
  Cl\'{e}ment, Auradou, and Ippolito]{MinoClement2018EColi}
G.~Mi\~{n}o, M.~Baabour, R.~Chertcoff, G.~Gutkind, E.~Cl\'{e}ment, H.~Auradou
  and I.~Ippolito, \emph{Adv. Microbiol.}, 2018, \textbf{8}, 451--464\relax
\mciteBstWouldAddEndPuncttrue
\mciteSetBstMidEndSepPunct{\mcitedefaultmidpunct}
{\mcitedefaultendpunct}{\mcitedefaultseppunct}\relax
\EndOfBibitem
\bibitem[Caprini and Marconi Marini~Bettolo(2018)]{caprini2019active}
L.~Caprini and U.~Marconi Marini~Bettolo, \emph{Soft Matter}, 2018,
  \textbf{14}, 9044--9054\relax
\mciteBstWouldAddEndPuncttrue
\mciteSetBstMidEndSepPunct{\mcitedefaultmidpunct}
{\mcitedefaultendpunct}{\mcitedefaultseppunct}\relax
\EndOfBibitem
\bibitem[Wensink and L{\"o}wen(2008)]{wensink2008aggregation}
H.~Wensink and H.~L{\"o}wen, \emph{Phys. Rev. E}, 2008, \textbf{78},
  031409\relax
\mciteBstWouldAddEndPuncttrue
\mciteSetBstMidEndSepPunct{\mcitedefaultmidpunct}
{\mcitedefaultendpunct}{\mcitedefaultseppunct}\relax
\EndOfBibitem
\bibitem[Kaiser \emph{et~al.}(2012)Kaiser, Wensink, and
  L{\"o}wen]{kaiser2012capture}
A.~Kaiser, H.~Wensink and H.~L{\"o}wen, \emph{Phys. Rev. Lett.}, 2012,
  \textbf{108}, 268307\relax
\mciteBstWouldAddEndPuncttrue
\mciteSetBstMidEndSepPunct{\mcitedefaultmidpunct}
{\mcitedefaultendpunct}{\mcitedefaultseppunct}\relax
\EndOfBibitem
\bibitem[Fily \emph{et~al.}(2014)Fily, Baskaran, and Hagan]{fily2014dynamics}
Y.~Fily, A.~Baskaran and M.~Hagan, \emph{Soft Matter}, 2014, \textbf{10},
  5609--5617\relax
\mciteBstWouldAddEndPuncttrue
\mciteSetBstMidEndSepPunct{\mcitedefaultmidpunct}
{\mcitedefaultendpunct}{\mcitedefaultseppunct}\relax
\EndOfBibitem
\bibitem[Elgeti and Gompper(2015)]{elgeti2015run}
J.~Elgeti and G.~Gompper, \emph{Europhys. Lett.}, 2015, \textbf{109},
  58003\relax
\mciteBstWouldAddEndPuncttrue
\mciteSetBstMidEndSepPunct{\mcitedefaultmidpunct}
{\mcitedefaultendpunct}{\mcitedefaultseppunct}\relax
\EndOfBibitem
\bibitem[J. and G.(2013)]{Elgeti_2013}
E.~J. and G.~G., \emph{Europhys. Lett.}, 2013, \textbf{101}, 48003\relax
\mciteBstWouldAddEndPuncttrue
\mciteSetBstMidEndSepPunct{\mcitedefaultmidpunct}
{\mcitedefaultendpunct}{\mcitedefaultseppunct}\relax
\EndOfBibitem
\bibitem[Fodor \emph{et~al.}(2016)Fodor, Nardini, Cates, Tailleur, Visco, and
  van Wijland]{fodor2016far}
E.~Fodor, C.~Nardini, M.~Cates, J.~Tailleur, P.~Visco and F.~van Wijland,
  \emph{Phys. Rev. Lett.}, 2016, \textbf{117}, 038103\relax
\mciteBstWouldAddEndPuncttrue
\mciteSetBstMidEndSepPunct{\mcitedefaultmidpunct}
{\mcitedefaultendpunct}{\mcitedefaultseppunct}\relax
\EndOfBibitem
\bibitem[Marconi Marini~Bettolo \emph{et~al.}(2017)Marconi Marini~Bettolo,
  Puglisi, and Maggi]{marconi2017heat}
U.~Marconi Marini~Bettolo, A.~Puglisi and C.~Maggi, \emph{Sci. Rep.}, 2017,
  \textbf{7}, 46496\relax
\mciteBstWouldAddEndPuncttrue
\mciteSetBstMidEndSepPunct{\mcitedefaultmidpunct}
{\mcitedefaultendpunct}{\mcitedefaultseppunct}\relax
\EndOfBibitem
\bibitem[Caprini \emph{et~al.}(2018)Caprini, Marconi Marini~Bettolo, and
  Vulpiani]{caprini2018linear}
L.~Caprini, U.~Marconi Marini~Bettolo and A.~Vulpiani, \emph{J. Stat. Mech.:
  Theory and Exp.}, 2018, \textbf{2018}, 033203\relax
\mciteBstWouldAddEndPuncttrue
\mciteSetBstMidEndSepPunct{\mcitedefaultmidpunct}
{\mcitedefaultendpunct}{\mcitedefaultseppunct}\relax
\EndOfBibitem
\bibitem[Takatori \emph{et~al.}(2016)Takatori, De~Dier, Vermant, and
  Brady]{takatori2016acoustic}
S.~C. Takatori, R.~De~Dier, J.~Vermant and J.~F. Brady, \emph{Nature Commun.},
  2016, \textbf{7}, 10694\relax
\mciteBstWouldAddEndPuncttrue
\mciteSetBstMidEndSepPunct{\mcitedefaultmidpunct}
{\mcitedefaultendpunct}{\mcitedefaultseppunct}\relax
\EndOfBibitem
\bibitem[Caprini \emph{et~al.}(2019)Caprini, Marconi Marini~Bettolo, and
  Puglisi]{caprini2019activity}
L.~Caprini, U.~Marconi Marini~Bettolo and A.~Puglisi, \emph{Sci. Rep.}, 2019,
  \textbf{9}, 1386\relax
\mciteBstWouldAddEndPuncttrue
\mciteSetBstMidEndSepPunct{\mcitedefaultmidpunct}
{\mcitedefaultendpunct}{\mcitedefaultseppunct}\relax
\EndOfBibitem
\bibitem[Caprini \emph{et~al.}(2019)Caprini, Marini Bettolo~Marconi, Puglisi,
  and Vulpiani]{caprini2018activeescape}
L.~Caprini, U.~Marini Bettolo~Marconi, A.~Puglisi and A.~Vulpiani, \emph{J.
  Chem. Phys.}, 2019, \textbf{150}, 024902\relax
\mciteBstWouldAddEndPuncttrue
\mciteSetBstMidEndSepPunct{\mcitedefaultmidpunct}
{\mcitedefaultendpunct}{\mcitedefaultseppunct}\relax
\EndOfBibitem
\bibitem[Fily and Marchetti(2012)]{fily2012athermal}
Y.~Fily and M.~Marchetti, \emph{Phys. Rev. Lett.}, 2012, \textbf{108},
  235702\relax
\mciteBstWouldAddEndPuncttrue
\mciteSetBstMidEndSepPunct{\mcitedefaultmidpunct}
{\mcitedefaultendpunct}{\mcitedefaultseppunct}\relax
\EndOfBibitem
\bibitem[Buttinoni \emph{et~al.}(2013)Buttinoni, Bialk{\'e}, K{\"u}mmel,
  L{\"o}wen, Bechinger, and Speck]{buttinoni2013dynamical}
I.~Buttinoni, J.~Bialk{\'e}, F.~K{\"u}mmel, H.~L{\"o}wen, C.~Bechinger and
  T.~Speck, \emph{Phys. Rev. Lett.}, 2013, \textbf{110}, 238301\relax
\mciteBstWouldAddEndPuncttrue
\mciteSetBstMidEndSepPunct{\mcitedefaultmidpunct}
{\mcitedefaultendpunct}{\mcitedefaultseppunct}\relax
\EndOfBibitem
\bibitem[Bialk{\'e} \emph{et~al.}(2015)Bialk{\'e}, Speck, and
  L{\"o}wen]{bialke2015active}
J.~Bialk{\'e}, T.~Speck and H.~L{\"o}wen, \emph{J. Non-Cryst. Solids}, 2015,
  \textbf{407}, 367--375\relax
\mciteBstWouldAddEndPuncttrue
\mciteSetBstMidEndSepPunct{\mcitedefaultmidpunct}
{\mcitedefaultendpunct}{\mcitedefaultseppunct}\relax
\EndOfBibitem
\bibitem[Cates and Tailleur(2015)]{cates2015motility}
M.~E. Cates and J.~Tailleur, \emph{Annu. Rev. Condens. Matter Phys.}, 2015,
  \textbf{6}, 219--244\relax
\mciteBstWouldAddEndPuncttrue
\mciteSetBstMidEndSepPunct{\mcitedefaultmidpunct}
{\mcitedefaultendpunct}{\mcitedefaultseppunct}\relax
\EndOfBibitem
\bibitem[Speck(2016)]{speck2016collective}
T.~Speck, \emph{The European Physical Journal Special Topics}, 2016,
  \textbf{225}, 2287--2299\relax
\mciteBstWouldAddEndPuncttrue
\mciteSetBstMidEndSepPunct{\mcitedefaultmidpunct}
{\mcitedefaultendpunct}{\mcitedefaultseppunct}\relax
\EndOfBibitem
\bibitem[Tjhung \emph{et~al.}(2018)Tjhung, Nardini, and
  Cates]{tjhung2018cluster}
E.~Tjhung, C.~Nardini and M.~Cates, \emph{Phys. Rev. X}, 2018, \textbf{8},
  031080\relax
\mciteBstWouldAddEndPuncttrue
\mciteSetBstMidEndSepPunct{\mcitedefaultmidpunct}
{\mcitedefaultendpunct}{\mcitedefaultseppunct}\relax
\EndOfBibitem
\bibitem[Digregorio \emph{et~al.}(2018)Digregorio, Levis, Suma, Cugliandolo,
  Gonnella, and Pagonabarraga]{digregorio2018full}
P.~Digregorio, D.~Levis, A.~Suma, L.~F. Cugliandolo, G.~Gonnella and
  I.~Pagonabarraga, \emph{Phys. Rev. Lett.}, 2018, \textbf{121}, 098003\relax
\mciteBstWouldAddEndPuncttrue
\mciteSetBstMidEndSepPunct{\mcitedefaultmidpunct}
{\mcitedefaultendpunct}{\mcitedefaultseppunct}\relax
\EndOfBibitem
\bibitem[Ribet and Cossart(2015)]{ribet2015bacterial}
D.~Ribet and P.~Cossart, \emph{Microbes and Infection}, 2015, \textbf{17},
  173--183\relax
\mciteBstWouldAddEndPuncttrue
\mciteSetBstMidEndSepPunct{\mcitedefaultmidpunct}
{\mcitedefaultendpunct}{\mcitedefaultseppunct}\relax
\EndOfBibitem
\bibitem[Malakar \emph{et~al.}(2018)Malakar, Jemseena, Kundu, Kumar,
  Sabhapandit, Majumdar, Redner, and Dhar]{malakar2018steady}
K.~Malakar, V.~Jemseena, A.~Kundu, K.~Kumar, S.~Sabhapandit, S.~Majumdar,
  S.~Redner and A.~Dhar, \emph{J. Stat. Mech. Theory Exp.}, 2018,
  \textbf{2018}, 043215\relax
\mciteBstWouldAddEndPuncttrue
\mciteSetBstMidEndSepPunct{\mcitedefaultmidpunct}
{\mcitedefaultendpunct}{\mcitedefaultseppunct}\relax
\EndOfBibitem
\bibitem[Weiss(2002)]{weiss2002some}
G.~H. Weiss, \emph{Physica A}, 2002, \textbf{311}, 381--410\relax
\mciteBstWouldAddEndPuncttrue
\mciteSetBstMidEndSepPunct{\mcitedefaultmidpunct}
{\mcitedefaultendpunct}{\mcitedefaultseppunct}\relax
\EndOfBibitem
\bibitem[Angelani \emph{et~al.}(2014)Angelani, Di~Leonardo, and
  Paoluzzi]{angelani2014first}
L.~Angelani, R.~Di~Leonardo and M.~Paoluzzi, \emph{Eur. Phys. J. E}, 2014,
  \textbf{37}, 59\relax
\mciteBstWouldAddEndPuncttrue
\mciteSetBstMidEndSepPunct{\mcitedefaultmidpunct}
{\mcitedefaultendpunct}{\mcitedefaultseppunct}\relax
\EndOfBibitem
\bibitem[Scacchi and Sharma(2018)]{scacchi2018mean}
A.~Scacchi and A.~Sharma, \emph{Mol. Phys.}, 2018, \textbf{116}, 460--464\relax
\mciteBstWouldAddEndPuncttrue
\mciteSetBstMidEndSepPunct{\mcitedefaultmidpunct}
{\mcitedefaultendpunct}{\mcitedefaultseppunct}\relax
\EndOfBibitem
\bibitem[Malgaretti and Stark(2017)]{malgaretti17}
P.~Malgaretti and H.~Stark, \emph{J. Chem. Phys.}, 2017, \textbf{146},
  174901\relax
\mciteBstWouldAddEndPuncttrue
\mciteSetBstMidEndSepPunct{\mcitedefaultmidpunct}
{\mcitedefaultendpunct}{\mcitedefaultseppunct}\relax
\EndOfBibitem
\bibitem[chun Wu \emph{et~al.}(2015)chun Wu, Chen, and quan Ai]{cinesi}
J.~chun Wu, Q.~Chen and B.~quan Ai, \emph{Phys. Lett. A}, 2015, \textbf{379},
  3025 -- 3028\relax
\mciteBstWouldAddEndPuncttrue
\mciteSetBstMidEndSepPunct{\mcitedefaultmidpunct}
{\mcitedefaultendpunct}{\mcitedefaultseppunct}\relax
\EndOfBibitem
\bibitem[Jacobs(1935)]{jacobsBook}
M.~H. Jacobs, \emph{Diffusion processes}, Springer, 1935, pp. 1--145\relax
\mciteBstWouldAddEndPuncttrue
\mciteSetBstMidEndSepPunct{\mcitedefaultmidpunct}
{\mcitedefaultendpunct}{\mcitedefaultseppunct}\relax
\EndOfBibitem
\bibitem[Zwanzig(1992)]{zwanzig}
R.~Zwanzig, \emph{J. Phys. Chem.}, 1992, \textbf{96}, 3926--3930\relax
\mciteBstWouldAddEndPuncttrue
\mciteSetBstMidEndSepPunct{\mcitedefaultmidpunct}
{\mcitedefaultendpunct}{\mcitedefaultseppunct}\relax
\EndOfBibitem
\bibitem[Reguera and Rub\'{\i}(2001)]{regueraFJ}
D.~Reguera and J.~M. Rub\'{\i}, \emph{Phys. Rev. E}, 2001, \textbf{64},
  061106\relax
\mciteBstWouldAddEndPuncttrue
\mciteSetBstMidEndSepPunct{\mcitedefaultmidpunct}
{\mcitedefaultendpunct}{\mcitedefaultseppunct}\relax
\EndOfBibitem
\bibitem[Burada \emph{et~al.}()Burada, Hänggi, Marchesoni, Schmid, and
  Talkner]{burada_Rev}
P.~Burada, P.~Hänggi, F.~Marchesoni, G.~Schmid and P.~Talkner,
  \emph{ChemPhysChem}, \textbf{10}, 45--54\relax
\mciteBstWouldAddEndPuncttrue
\mciteSetBstMidEndSepPunct{\mcitedefaultmidpunct}
{\mcitedefaultendpunct}{\mcitedefaultseppunct}\relax
\EndOfBibitem
\bibitem[Sandoval and Dagdug(2014)]{dagdug14}
M.~Sandoval and L.~Dagdug, \emph{Phys. Rev. E}, 2014, \textbf{90}, 062711\relax
\mciteBstWouldAddEndPuncttrue
\mciteSetBstMidEndSepPunct{\mcitedefaultmidpunct}
{\mcitedefaultendpunct}{\mcitedefaultseppunct}\relax
\EndOfBibitem
\bibitem[Sevilla \emph{et~al.}(2019)Sevilla, Arzola, and
  Cital]{sevilla2019stationary}
F.~J. Sevilla, A.~V. Arzola and E.~P. Cital, \emph{Phys. Rev. E}, 2019,
  \textbf{99}, 012145\relax
\mciteBstWouldAddEndPuncttrue
\mciteSetBstMidEndSepPunct{\mcitedefaultmidpunct}
{\mcitedefaultendpunct}{\mcitedefaultseppunct}\relax
\EndOfBibitem
\bibitem[Fily(2018)]{fily2018self}
Y.~Fily, \emph{arXiv preprint arXiv:1812.05698}, 2018\relax
\mciteBstWouldAddEndPuncttrue
\mciteSetBstMidEndSepPunct{\mcitedefaultmidpunct}
{\mcitedefaultendpunct}{\mcitedefaultseppunct}\relax
\EndOfBibitem
\bibitem[Sharma \emph{et~al.}(2017)Sharma, Wittmann, and
  Brader]{sharma2017escape}
A.~Sharma, R.~Wittmann and J.~Brader, \emph{Phys. Rev. E}, 2017, \textbf{95},
  012115\relax
\mciteBstWouldAddEndPuncttrue
\mciteSetBstMidEndSepPunct{\mcitedefaultmidpunct}
{\mcitedefaultendpunct}{\mcitedefaultseppunct}\relax
\EndOfBibitem
\bibitem[Bechinger \emph{et~al.}(2013)Bechinger, Sciortino, and
  Ziherl]{bechinger2013physics}
C.~Bechinger, F.~Sciortino and P.~Ziherl, \emph{Physics of complex colloids},
  IOS Press, 2013, vol. 184\relax
\mciteBstWouldAddEndPuncttrue
\mciteSetBstMidEndSepPunct{\mcitedefaultmidpunct}
{\mcitedefaultendpunct}{\mcitedefaultseppunct}\relax
\EndOfBibitem
\bibitem[Romanczuk and Schimansky-Geier(2011)]{romanczuk2011brownian}
P.~Romanczuk and L.~Schimansky-Geier, \emph{Phys. Rev. Lett.}, 2011,
  \textbf{106}, 230601\relax
\mciteBstWouldAddEndPuncttrue
\mciteSetBstMidEndSepPunct{\mcitedefaultmidpunct}
{\mcitedefaultendpunct}{\mcitedefaultseppunct}\relax
\EndOfBibitem
\bibitem[Toral and Colet(2014)]{toral2014stochastic}
R.~Toral and P.~Colet, \emph{Stochastic numerical methods: an introduction for
  students and scientists}, John Wiley \& Sons, 2014\relax
\mciteBstWouldAddEndPuncttrue
\mciteSetBstMidEndSepPunct{\mcitedefaultmidpunct}
{\mcitedefaultendpunct}{\mcitedefaultseppunct}\relax
\EndOfBibitem
\bibitem[Maggi \emph{et~al.}(2015)Maggi, Marconi, Gnan, and
  Di~Leonardo]{maggi2015multidimensional}
C.~Maggi, U.~M.~B. Marconi, N.~Gnan and R.~Di~Leonardo, \emph{Sci. Rep.}, 2015,
  \textbf{5}, 10742\relax
\mciteBstWouldAddEndPuncttrue
\mciteSetBstMidEndSepPunct{\mcitedefaultmidpunct}
{\mcitedefaultendpunct}{\mcitedefaultseppunct}\relax
\EndOfBibitem
\bibitem[Wagner \emph{et~al.}(2017)Wagner, Hagan, and
  Baskaran]{wagner2017steady}
C.~G. Wagner, M.~F. Hagan and A.~Baskaran, \emph{J. Stat. Mech.: Theory and
  Exp.}, 2017, \textbf{2017}, 043203\relax
\mciteBstWouldAddEndPuncttrue
\mciteSetBstMidEndSepPunct{\mcitedefaultmidpunct}
{\mcitedefaultendpunct}{\mcitedefaultseppunct}\relax
\EndOfBibitem
\bibitem[Angelani(2017)]{angelani2017confined}
L.~Angelani, \emph{J. Phys. A: Math. and Theor.}, 2017, \textbf{50},
  325601\relax
\mciteBstWouldAddEndPuncttrue
\mciteSetBstMidEndSepPunct{\mcitedefaultmidpunct}
{\mcitedefaultendpunct}{\mcitedefaultseppunct}\relax
\EndOfBibitem
\bibitem[Lee(2013)]{lee2013active}
C.~Lee, \emph{New J Phys.}, 2013, \textbf{15}, 055007\relax
\mciteBstWouldAddEndPuncttrue
\mciteSetBstMidEndSepPunct{\mcitedefaultmidpunct}
{\mcitedefaultendpunct}{\mcitedefaultseppunct}\relax
\EndOfBibitem
\bibitem[Winkler \emph{et~al.}(2015)Winkler, Wysocki, and
  Gompper]{winkler2015virial}
R.~G. Winkler, A.~Wysocki and G.~Gompper, \emph{Soft Matter}, 2015,
  \textbf{11}, 6680--6691\relax
\mciteBstWouldAddEndPuncttrue
\mciteSetBstMidEndSepPunct{\mcitedefaultmidpunct}
{\mcitedefaultendpunct}{\mcitedefaultseppunct}\relax
\EndOfBibitem
\bibitem[Kurzthaler \emph{et~al.}(2016)Kurzthaler, Leitmann, and
  Franosch]{kurzthaler2016intermediate}
C.~Kurzthaler, S.~Leitmann and T.~Franosch, \emph{Sci. Rep.}, 2016, \textbf{6},
  36702\relax
\mciteBstWouldAddEndPuncttrue
\mciteSetBstMidEndSepPunct{\mcitedefaultmidpunct}
{\mcitedefaultendpunct}{\mcitedefaultseppunct}\relax
\EndOfBibitem
\bibitem[Burada \emph{et~al.}(2007)Burada, Schmid, Reguera, Rub\'{\i}, and
  H\"anggi]{buradaTEST}
P.~S. Burada, G.~Schmid, D.~Reguera, J.~M. Rub\'{\i} and P.~H\"anggi,
  \emph{Phys. Rev. E}, 2007, \textbf{75}, 051111(8)\relax
\mciteBstWouldAddEndPuncttrue
\mciteSetBstMidEndSepPunct{\mcitedefaultmidpunct}
{\mcitedefaultendpunct}{\mcitedefaultseppunct}\relax
\EndOfBibitem
\bibitem[Forte \emph{et~al.}(2014)Forte, Cecconi, and Vulpiani]{fortePRE}
G.~Forte, F.~Cecconi and A.~Vulpiani, \emph{Phys. Rev. E}, 2014, \textbf{90},
  062110(10)\relax
\mciteBstWouldAddEndPuncttrue
\mciteSetBstMidEndSepPunct{\mcitedefaultmidpunct}
{\mcitedefaultendpunct}{\mcitedefaultseppunct}\relax
\EndOfBibitem
\bibitem[Arfken and Weber(2001)]{arfken2001}
G.~Arfken and H.~Weber, \emph{Mathematical methods for physicists}, 2001\relax
\mciteBstWouldAddEndPuncttrue
\mciteSetBstMidEndSepPunct{\mcitedefaultmidpunct}
{\mcitedefaultendpunct}{\mcitedefaultseppunct}\relax
\EndOfBibitem
\bibitem[Ghosh \emph{et~al.}(2013)Ghosh, Misko, Marchesoni, and
  Nori]{Janus_Ratchet}
P.~K. Ghosh, V.~R. Misko, F.~Marchesoni and F.~Nori, \emph{Phys. Rev. Lett.},
  2013, \textbf{110}, 268301(5)\relax
\mciteBstWouldAddEndPuncttrue
\mciteSetBstMidEndSepPunct{\mcitedefaultmidpunct}
{\mcitedefaultendpunct}{\mcitedefaultseppunct}\relax
\EndOfBibitem
\bibitem[Gardiner(2009)]{Gardiner}
C.~Gardiner, \emph{Handbook of Stochastic: For the Natural and Social
  Sciences}, Springer, Berlin, 2009, vol.~4\relax
\mciteBstWouldAddEndPuncttrue
\mciteSetBstMidEndSepPunct{\mcitedefaultmidpunct}
{\mcitedefaultendpunct}{\mcitedefaultseppunct}\relax
\EndOfBibitem
\end{mcitethebibliography}

\providecommand*{\mcitethebibliography}{\thebibliography}
\csname @ifundefined\endcsname{endmcitethebibliography}
{\let\endmcitethebibliography\endthebibliography}{}

\end{document}